\documentclass[acmsmall,screen]{acmart}
\AtBeginDocument{%
  }

\setcopyright{acmlicensed}
\copyrightyear{2027}
\acmYear{2027}
\acmDOI{XXXXXXX.XXXXXXX}
\acmISBN{978-1-4503-XXXX-X/2018/06}

\usepackage{listings}
\usepackage{xcolor}
\usepackage{makecell}
\usepackage{tabularx}
\usepackage{fvextra}
\usepackage{booktabs}
\usepackage{enumitem}
\usepackage{array}
\usepackage[table]{xcolor}
\usepackage{tcolorbox}
\usepackage{listings}
\usepackage{amsmath}
\usepackage{subcaption}
\usepackage{tikz}
\tcbuselibrary{skins, breakable}
\usepackage{setspace}
\usepackage{multirow}
\usepackage{pifont}
\usepackage{lstlinebgrd}
\usepackage{amsthm}
\newsavebox{\mycodebox}
\usepackage{threeparttable} 
\usepackage{url}
\usetikzlibrary{positioning}
\definecolor{codegreen}{rgb}{0,0.6,0}
\definecolor{codegray}{rgb}{0.5,0.5,0.5}
\definecolor{codepurple}{rgb}{0.58,0,0.82}
\definecolor{backcolour}{rgb}{0.95,0.95,0.92}
\definecolor{userbubble}{RGB}{0, 122, 255}   
\definecolor{llmbubble}{RGB}{229, 229, 234}  
\definecolor{usertext}{RGB}{255, 255, 255}
\definecolor{llmtext}{RGB}{0, 0, 0}
\definecolor{chatbg}{RGB}{245, 245, 245}
\definecolor{defbg}{RGB}{245,248,255}
\definecolor{defframe}{RGB}{100,130,200}
\definecolor{deftitle}{RGB}{30,60,140}

\definecolor{codeblue}{RGB}{240,246,255}
\definecolor{cotgray}{RGB}{245,245,245}
\definecolor{veragreen}{RGB}{240,248,240}
\definecolor{failred}{RGB}{220,50,47}
\definecolor{passgreen}{RGB}{42,130,42}
\definecolor{annotblue}{RGB}{30,80,180}
\definecolor{llmgray}{gray}{0.92}

\definecolor{prored}{RGB}{190, 30, 45}

\newcommand{\red}[1]{{\color{prored}#1}}
\newcommand{\boldparagraph}[1]{\vskip 0.05in\noindent\textbf{#1.}}

\newif \ifnotes
\notesfalse
\ifnotes
\newcommand{\notes}[2]{{\bf\textsf{\textcolor{#1}{#2}}}} 
\else \newcommand{\notes}[2]{} 
\fi
\newcommand{\limin}[1]{\notes{magenta}{Limin says: #1}}
\newcommand{\ronghao}[1]{\notes{orange}{Ronghao says: #1}}

\newcommand{\caroline}[1]{\notes{blue}{Caroline: #1}}

\begin{document}

\title{Correct Verdicts, Flawed Reasoning: Structured Auditing of LLM-based Vulnerability Reasoning}

\author{Boyue Caroline Hu}
\email{boyuecah@andrew.cmu.edu}
\affiliation{%
  \institution{Carnegie Mellon University}
  \city{Pittsburgh}
  \state{Pennsilvanya}
  \country{USA}
}

\author{Kaivalya Ahir}
\email{kahir@andrew.cmu.edu}
\affiliation{%
  \institution{Carnegie Mellon University}
  \city{Pittsburgh}
  \state{Pennsilvanya}
  \country{USA}
}
\author{Ronghao Ni}
\email{ronghaon@andrew.cmu.edu}
\affiliation{%
  \institution{Carnegie Mellon University}
  \city{Pittsburgh}
  \state{Pennsilvanya}
  \country{USA}
}

\author{Limin Jia}
\email{liminjia@andrew.cmu.edu}
\affiliation{%
  \institution{Carnegie Mellon University}
  \city{Pittsburgh}
  \state{Pennsilvanya}
  \country{USA}
}


\begin{abstract}
Large Language Models (LLMs) are increasingly deployed for automated software vulnerability analysis. Binary classification alone is often insufficient; practitioners need explanations to triage reported bugs and engineer patches. 
Standard practice relies on Chain-of-Thought (CoT) prompting to elicit such explanations. However, because CoT represents reasoning as free-form natural language, models can obscure logical leaps, hallucinated execution steps, and internal inconsistencies behind plausible prose, leading to \emph{correct verdict, flawed reasoning}, such as reaching the correct verdict via a non-existent execution path. Our manual audit reveals that approximately $60\%$ of correct vulnerability verdicts are accompanied by these fabricated or unverifiable claims. Free-form explanations allow reasoning errors to evade LLM-as-a-judge evaluation.

To better audit LLM reasoning, we present \textbf{V}ulnerability \textbf{E}xplanation \textbf{R}easoning \textbf{A}uditor (\textbf{VERA}), an automated framework for auditing whether the explanation adheres to the source code; and whether it contains logical contradiction. Rather than accepting free-form text, VERA first asks models to output a Structured Reasoning Record (SRR), encoding the information necessary for vulnerability reasoning into machine-readable fields, including tracked pointers, memory operations, and state transitions. A multi-stage judge then audits each SRR against eight well-defined reasoning failure modes. Most stages use deterministic syntactic checks, with LLM calls reserved only when semantic interpretation is required. Furthermore, the standardized schema of SRRs enables automated mutation testing  to benchmark reasoning judges at scale without human annotation. Our empirical evaluation demonstrates that reasoning flaws occur in correct verdicts just as frequently as in incorrect ones, and VERA exposes $87\%$ of reasoning errors that free-form LLM-as-judge systematically miss. VERA replaces blind trust in CoT prose with structured reasoning audits, filtering hallucinated vulnerability reports and pinpointing stage-specific failures.

\end{abstract}


 
\maketitle

\section{Introduction}
\label{sec:intro} 
Large Language Models (LLMs) are increasingly being adopted for automated vulnerability detection in software security~\cite{Ding2024PrimeVul,llm_security_survey,GPT_4_security,YU2026,Fang2024}. 
Trust in LLM-generated vulnerability assessments depends not only on the final verdict but also on the underlying reasoning, especially in security-critical settings~\cite{johnson2013,lu2026dagvul}. Explanations of where and why a vulnerability exists help developers and security analysts validate findings and make informed patching and auditing decisions.
However, LLMs may provide hallucinated explanations and justifications~\cite{hallucination_survey}. 
 A common LLM-based vulnerability analysis approach is to use unconstrained Chain-of-Thought (CoT) reasoning in natural language.
While flexible, free-form CoT allows models to produce plausible-sounding but fundamentally hallucinated justifications~\cite{lekssays2025llmxcpg,shortcut2020,PearceA0DK22,turpin2023unfaithful,lyu2023faithful}, a phenomenon we refer to as ``correct verdicts, flawed reasoning.''
%
%


\definecolor{codeblue}{RGB}{240,246,255}
\definecolor{cotgray}{RGB}{245,245,245}
\definecolor{veragreen}{RGB}{240,248,240}
\definecolor{failred}{RGB}{220,50,47}
\definecolor{passgreen}{RGB}{42,130,42}
\definecolor{annotblue}{RGB}{30,80,180}
\definecolor{warnorg}{RGB}{200,90,0}

\newcommand{\micro}{\fontsize{7}{6}\selectfont}
\newcommand{\microit}{\fontsize{7}{6}\selectfont\itshape}
\newcommand{\microbf}{\fontsize{7}{6}\selectfont\bfseries}

\lstdefinestyle{cstyle}{
  language=C, basicstyle=\ttfamily\fontsize{6}{6}\selectfont,
  keywordstyle=\bfseries\color{blue!70!black},
  commentstyle=\itshape\color{gray},
  numbers=left, numberstyle=\fontsize{6}{5}\selectfont\color{gray}, numbersep=4pt,
  xleftmargin=10pt,
  breaklines=true, frame=none,
  escapeinside={(*@}{@*)}, tabsize=2, showstringspaces=false,
}

\tcbset{
  panelbox/.style={
    enhanced, sharp corners, boxrule=0.4pt, colframe=gray!50,
    fonttitle=\bfseries\fontsize{6}{6}\selectfont, coltitle=black,
    top=1pt, bottom=1pt, left=2pt, right=2pt,
    toptitle=1pt, bottomtitle=0pt,
  },
  codepanel/.style={panelbox, colback=codeblue,  colbacktitle=blue!15!white},
  cotpanel/.style= {panelbox, colback=cotgray,   colbacktitle=gray!25},
  verapanel/.style={panelbox, colback=veragreen,  colbacktitle=green!15!white},
  defstrip/.style={
    enhanced, sharp corners, boxrule=0.4pt, colframe=gray!40,
    colback=gray!6, top=2pt, bottom=2pt, left=1pt, right=4pt,
    fontupper=\fontsize{5}{6}\selectfont,
  },
  stagebox/.style={
    enhanced, sharp corners, boxrule=0.3pt, colframe=gray!30,
    colback=white, top=1pt, bottom=0pt, left=2pt, right=2pt,
    fontupper=\fontsize{5}{6}\selectfont,
  },
  holisticpass/.style={
    enhanced, sharp corners, boxrule=0.4pt, colframe=passgreen!50,
    colback=passgreen!6, top=0pt, bottom=0pt, left=1pt, right=1pt,
    fontupper=\fontsize{5}{6}\selectfont,
  },
}

\newcommand{\annot}[1]{%
  \tikz[baseline=(x.base)]{\node[circle,draw=annotblue,fill=annotblue!15,
    inner sep=0.6pt,font=\bfseries\fontsize{4.5}{5}\selectfont,
    text=annotblue](x){#1};}%
}
\newcommand{\annotc}[1]{%
  \textcolor{annotblue}{\textbf{\fbox{\fontsize{4.5}{5}\selectfont#1}}}%
}
\newcommand{\xfail}{\textcolor{failred}{\textbf{\texttimes}}}
\newcommand{\xpass}{\textcolor{passgreen}{\checkmark}}
\newcommand{\sann}[1]{\textcolor{annotblue}{\micro\textit{$\triangleright$\,#1}}}

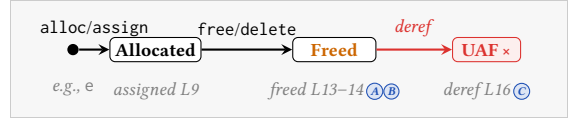
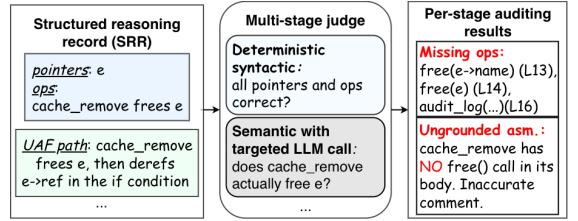
\begin{figure*}[t]
\centering\small
\setlength{\fboxsep}{0pt}

\begin{subfigure}[t]{\textwidth}
\includegraphics[width=\linewidth]{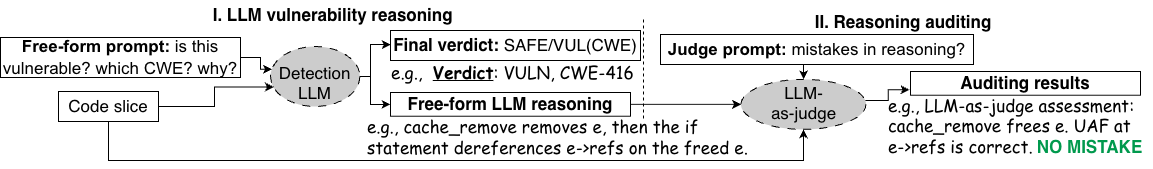}
\vspace{-13pt}
\caption{Chain-of-Thought (CoT) detection and LLM-as-judge baseline. Example on the right shows the LLM fabricates a cause via a stale comment; free-form prose makes the error invisible to the LLM-as-judge. }
\label{fig:overview-cot}
\end{subfigure}
\hfill
\begin{subfigure}[t]{\textwidth}
\includegraphics[width=\linewidth]{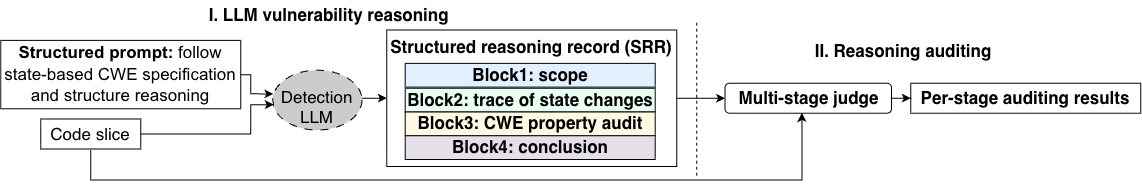}
\vspace{-13pt}
\caption{Overview of VERA. VERA asks the detection LLM to structure its reasoning into an SRR, then audits it with a multi-stage judge, producing per-stage auditing results. \ronghao{Maybe better if we can have a vectorized format such as PDF? I'm thinking whether we can make a and b more contrastive to show a is what existing works explore and is bad while b is ours and is good? Add some vertical texts on the left side of the figures?}}
\label{fig:overview-vera}
\end{subfigure}

\vspace{6pt}

\begin{minipage}[t]{0.45\textwidth}
\vspace{0pt}

\begin{subfigure}[t]{\linewidth}
\begin{tcolorbox}[codepanel,
  title={\texttt{evict\_entry(cache\_t *cache, int key)}\enspace
    {\color{annotblue}\micro\textit{$\triangleright$\,=\, variable state}}}]
\begin{lstlisting}[style=cstyle]
typedef struct {
  char *name;  int refs;
} entry_t;

void cache_remove(cache_t *cache, int key) {(*@\textcolor{gray}{\microit{/*removes key from index only*/}}@*)
  ... }(*@\textcolor{red!60!black}{\microbf{~~\%Note: not freeing e\%}}@*)
  
void evict_entry(cache_t *cache, int key) {
  entry_t *e = cache_lookup(cache, key);
  if (!e) return;
  cache_remove(cache, key); /* removes entry e (*@\textcolor{red!60!black}{\microbf{~~\%Note: stale comment\%}}@*) */
  if (--e->refs == 0) {
    free(e->name); (*@\annot{A}\sann{e->name: \textcolor{orange!80!black}{Freed}}@*)
    free(e);  (*@\annot{B}\sann{e: \textcolor{orange!80!black}{Freed}}@*)
  }               
  audit_log("evicted \%d: \%s", key, e->name); (*@\annot{C}\sann{\textcolor{failred}{UAF: e->name is \textcolor{orange!80!black}{Freed}}}@*)
  }
\end{lstlisting}
\end{tcolorbox}
\vspace{-8pt}
\caption{When \texttt{e->refs} reaches zero, \annot{A}\,\texttt{e->name}
  and \annot{B}\,\texttt{e} are freed. Then \texttt{audit\_log}
  dereferences \texttt{e->name} in the
  \textcolor{orange!80!black}{\textbf{Freed}} state, causing the UAF.}
\label{fig:motivating:code}
\end{subfigure}

\end{minipage}
\hfill
\begin{minipage}[t]{0.53\textwidth}
\vspace{0pt}
\begin{subfigure}[t]{\linewidth}
\begin{tcolorbox}[defstrip]
\centering
\begin{tikzpicture}[
  font=\fontsize{8}{6}\selectfont,
  state/.style={draw, rounded corners=2pt, fill=white,
                inner sep=2pt, minimum width=1.1cm, minimum height=0.32cm,
                font=\fontsize{7}{6}\selectfont\bfseries},
  err/.style={draw=failred, rounded corners=2pt, fill=failred!8,
              inner sep=2pt, minimum width=0.9cm, minimum height=0.32cm,
              font=\fontsize{7}{6}\selectfont\bfseries, text=failred},
  arr/.style={->, >=stealth, thick, font=\fontsize{7}{5}\selectfont},
  redarr/.style={->, >=stealth, thick, color=failred,
                 font=\fontsize{7}{5}\selectfont},
  eg/.style={font=\fontsize{7}{5}\selectfont\itshape, text=gray, anchor=north},
]
  \node[circle, fill=black, inner sep=1.5pt]          (start) {};
  \node[state, right=0.4cm of start]                  (alloc) {\textbf{Allocated}};
  \node[state, right=1.2cm of alloc,
        text=orange!80!black]                          (freed) {\textbf{Freed}};
  \node[err,   right=1.0cm of freed]                  (uaf)   {\textbf{UAF}\,\xfail};

  \draw[arr] (start) -- node[above, yshift=3pt]{\texttt{alloc}/\texttt{assign}} (alloc);
  \draw[arr] (alloc) -- node[above, yshift=3pt]{\texttt{free}/\texttt{delete}}  (freed);
  \draw[redarr] (freed) -- node[above, yshift=3pt]{\textit{deref}}              (uaf);

  \coordinate (egrow) at ([yshift=-0.2cm]alloc.south);
  \node[eg] at (start |- egrow) {e.g., \texttt{e}};
  \node[eg] at (alloc |- egrow) {assigned L9};
  \node[eg] at (freed |- egrow) {freed L13--14\,\annot{A}\annot{B}};
  \node[eg] at (uaf   |- egrow) {deref L16\,\annot{C}};

\end{tikzpicture}
\end{tcolorbox}
\vspace{-5pt}
\caption{State-based specification of CWE-416 Use-After-Free (UAF): accessing pointers in the \texttt{Freed} state. }
\label{fig:motivating:cwe}
\end{subfigure}
\vspace{4pt}

\begin{subfigure}[t]{\linewidth}
\includegraphics[width=\linewidth]{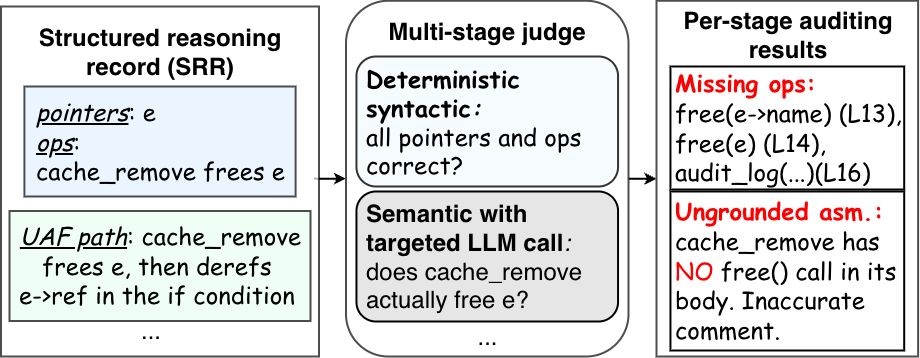}
\vspace{-7pt}
\caption{Example of reasoning auditing with VERA. Structuring reasoning into an auditable SRR enables judges to detect missed memory operations and false assumptions. 
}
\label{fig:motivating:vera-ex}
\end{subfigure}

\end{minipage}
\vspace{-5pt}
\caption{Overview of VERA with a motivating example of \emph{correct verdict, flawed reasoning}. }
\vspace{-10pt}
\label{fig:motivating}
\end{figure*}

For example, the program in Fig.~\ref{fig:motivating:code} has a Use-After-Free (UAF) vulnerability (CWE-416~\cite{cwe416}). 
A common way to explain the vulnerability is that when the reference count is zero (L12), the pointers \texttt{e->name} and \texttt{e} are freed on lines L13,~14, then the dereferencing \texttt{e->name} in \texttt{audit\_log} (L16) leads to UAF.  
When prompted to identify and explain the vulnerability in this program, as illustrated in Fig.~\ref{fig:overview-cot}, LLM via CoT correctly identified UAF, but its reasoning falsely assumes \texttt{cache\_remove} frees \texttt{e} (L11) then dereferencing \texttt{e->refs} (L12) is the UAF. This error violates basic \emph{code adherence} as the reasoning was based on the stale comment (L11) rather than actual implementation. Though not shown in the example, reasoning could also contain logical contradictions, such as explicitly identifying a vulnerability step-by-step only to conclude with a ``Safe'' final verdict. 

Our manual review of free-form CoT reasoning generated by four state-of-the-art LLMs revealed that LLMs frequently provide explanations with hallucinated external knowledge ($55\text{-}61\%$ factually incorrect) or unverifiable code assumptions ($19\text{-}34\%$) (more details in Sec.~\ref{sec:motivation}). 
Relying on flawed explanations in production carries significant risk: engineers acting on hallucinated logic may write broken patches, while high benchmark accuracy conceals critical model failure modes (e.g., missing the actual UAF in Fig.~\ref{fig:motivating:code}). Addressing this requires a framework that checks 
whether underlying reasoning by the LLMs is both grounded in facts about the source code and are logically consistent.

The primary challenge of auditing LLM vulnerability reasoning is that free-form CoT reasoning lacks explicit structure and intermediate checkable steps, e.g., where pointers are freed and dereferenced. Furthermore, unlike algorithmic or boolean problems where intermediate logical steps unfold sequentially, checkable reasoning steps for vulnerability analysis need to be explicitly formalized and structured. 
Hence, recent efforts have turned to LLM-as-judge approaches~\cite{guo2026bugscope,zhu2026vulinstruct,lu2026dagvul} (Step~II of Fig.~\ref{fig:overview-cot}); however, evaluating unstructured natural language with another neural model is still prone to hallucinations and yields no verifiable evidences.
For instance, in Fig.~\ref{fig:overview-cot}, the LLM-as-judge is likewise misled by the stale comment and misses the error. Human experts can detect such issues, but their review does not scale.

Our core insight is that for vulnerability analysis, checkable intermediate reasoning steps naturally correspond to program state transitions and that a broad spectrum of vulnerability classes, from memory safety to resource lifecycle errors, can be specified as state sequences.
%
%
For instance, finding a UAF can be described as a process of finding operations that transition a pointer to \textit{Allocated} state and then to \textit{Freed}, and look for a dereference operation of any \textit{Freed} pointer. In Fig.~\ref{fig:motivating:code}, this sequence is \texttt{e->refs==0} (L12), then \texttt{e->name} and \texttt{e} are freed (L13,~14), and the dereferencing \texttt{e->name} in \texttt{audit\_log} (L16) leads to UAF (Fig.~\ref{fig:motivating:cwe}). 
By instructing the LLM to explicitly track and output these state changes, we transform natural language reasoning into a machine-readable schema that can be deterministically checked against the input source code. 

Building on this insight, we present \textbf{VERA} (\textbf{V}ulnerability \textbf{E}xplanation \textbf{R}easoning \textbf{A}uditor), a framework that constrains LLM explanations to structured, auditable reasoning traces and performs automatic audit. Fig.~\ref{fig:overview-vera} illustrates the architecture of VERA.
We focus on vulnerability reasoning by LLMs when given a slice of the project (e.g., key function definitions). This is widely favored in practice because it is computationally efficient, scales across massive codebases, and operates directly on source code without requiring a fully compilable build environment. 
Rather than relying on high-level CWE categories paired with free-form CoT prompts, we first specify vulnerabilities as sequences of state transitions. VERA then instructs the LLM to output a Structured Reasoning Record (SRR) that explicitly materializes checkable intermediate reasoning steps, including variable states, state-changing operations, and operational assumptions, into an inspectable format for auditing (Step~I in Fig.~\ref{fig:overview-vera}). In the example in Fig.~\ref{fig:motivating:vera-ex}, the SRR explicitly records the intermediate assumption about \texttt{cache\_remove} that it frees \texttt{e}, then the conclusion that \texttt{e->refs} is the site of the UAF that can later be detected during auditing.

Next, VERA performs \emph{reasoning auditing} by evaluating the SRR through a multi-stage judge to detect reasoning breakdowns across a comprehensive taxonomy of fundamental failure modes (Step~II). 
We define reasoning auditing as evaluating \emph{code adherence} and \emph{internal consistency}. The former refers to checking whether each claim in an LLM's reasoning chain is syntactically and semantically supported by the provided source code slice~\cite{guo2026bugscope} (\textit{e.g.}, whether variables named in the trace are actually declared). The later refers to verifying that intermediate trace steps remain logically coherent and strictly support the final verdict (\textit{e.g.}, ensuring the final vulnerable verdict does not contradict the trace). Note that neither check requires ground-truth reasoning annotations which do not exist in real-world applications. Existing vulnerability validation techniques are fundamentally unsuited for this task. Benchmark studies evaluate only final detection metrics, treating internal reasoning as a black box~\cite{bigvul,Ding2024PrimeVul,chen2023diversevul}. 
Traditional static tools~\cite{codeql,joern,semgrep2024} 
focus on rule-based detection rather than explanatory auditability, frequently flooding triage queues with high false-positive rates without explaining root causes. Dynamic validation methods (\textit{e.g.}, fuzzing~\cite{afl_plus,fuzz4all,pham2019smart} or sanitizer-based execution~\cite{asan,valgrind}) and LLM-based vulnerability validation~\cite{guo2026bugscope,zhu2026vulinstruct,lu2026dagvul} 
have a fundamentally different goal from reasoning auditing. They confirm whether the detected bug exists or executes at runtime rather than whether the explanatory reasoning is sound. 
%
To remain broadly applicable across diverse programming languages and vulnerability classes (CWEs) without requiring language-specific static analyzers or heavy compilation toolchains, VERA's judge relies primarily on deterministic AST parsing for syntactic checking, reserving targeted LLM calls strictly for higher-level semantic checks. By evaluating the SRR rather than unstructured text, VERA exposes both reasoning mistakes in the example: the omitted memory operations and the invalid assumption drawn from the stale comment.

In summary, this paper makes the following contributions. 1) \textbf{VERA: A Reasoning Audit Framework for Vulnerability Analysis.} VERA takes an initial step toward systematic LLM reasoning validation in software security by providing an automated framework to audit code adherence and internal consistency in vulnerability reasoning. 
2) \textbf{Automated Mutation Testing for Reasoning Judges.} 
Evaluating reasoning evaluators has traditionally required unscalable human inspection. By formalizing reasoning into explicit, machine-readable SRRs, VERA enables automated, ground-truth-free mutation testing of reasoning judges, providing a scalable framework to automatically benchmark judge precision and recall without human spot-checks.
\textbf{3) Empirical Findings \& Practical Impact:}
Our experiments using VERA reveal three critical insights. 
First, reasoning flaws occur in correct verdicts just as frequently as incorrect ones, models are right for the wrong reasons as often as they are simply wrong, making reasoning audit necessary. Second, false positives are systematic and structure-driven: models repeat identical reasoning mistakes on safe functions and their patched variants. Third, VERA's per-stage decomposition reveals code-structural patterns that drive specific reasoning failures.
By surfacing failures as explicit stage-level findings, VERA reduces manual effort to audit LLM reasoning chains for security analysts and provides LLM developers with actionable diagnostics for targeted fine-tuning and benchmark design.

The rest of this paper is organized as follows. Sec.~\ref{sec:motivation} motives VERA by demonstrating empirically that CoT reasoning is unreliable and unauditable. Sec.~\ref{sec:structure} introduces our state-based vulnerability specifications, reasoning failure taxonomy, and the Structured Reasoning Record (SRR) schema of VERA. Sec.~\ref{sec:judge} presents the multi-stage judge and mutation testing for judges. Sec.~\ref{sec:evaluation} reports evaluation results. Finally, we review related work in Sec.~\ref{sec:related_work} and conclude in Sec.~\ref{sec:conclusion}.

\section{Motivation: Free-form CoT Obscures Reasoning Mistakes}
\label{sec:motivation}

\begin{table}[t]
\centering
\small
\caption{Detection reliability and reasoning quality of free-form CoT (64 CWE-416 pairs, SVEN). }
\label{tab:motivation}
\begin{minipage}[t]{0.70\linewidth}
\vspace{-5pt}
  \resizebox{\linewidth}{!}{%
  \begin{tabular}{lcccccc}
  \toprule
  & \multicolumn{4}{c}{\textit{Verdict Correctness Metrics}} & \textit{Ext. Knowledge} & \textit{Reasoning} \\
  \cmidrule(lr){2-5} \cmidrule(lr){6-6} \cmidrule(lr){7-7}
  \textbf{Model}
    & \textbf{Binary Acc.}
    & \textbf{CWE recall}
    & \textbf{FP}
    & \textbf{Both V\%}\textsuperscript{a}
    & \textbf{Seen\textsuperscript{b} / Wrong\%}
    & \textbf{Vague\%}\textsuperscript{c} \\
  \midrule
  GPT-5.5~\cite{singh2026openaigpt5card}       & 60\% & 16\% & 49\% & 44\% &  2\% / --- & 19\% \\
  GPT-5.6-sol~\cite{singh2026openaigpt5card}   & 59\% & 13\% & 38\% & 33\% &  0\% / --- & 28\% \\
  Gemma-4-26b~\cite{google2026gemma}   & 46\% &  3\% & 74\% & 58\% &  0\% / --- & 22\% \\
  Sonnet-5~\cite{anthropic2026claude}      & 59\% & 13\% & 46\% & 42\% & 44\% / 61\% & 34\% \\
  \bottomrule
  \end{tabular}}
\end{minipage}%
\hfill
\begin{minipage}[t]{0.28\linewidth}
  \vspace{0pt}
  \scriptsize
  \textsuperscript{a}~Both received a \textsc{vulnerable} verdict.\\[4pt]
  \textsuperscript{b}~Cites external knowledge (e.g., known CVE/fix); \textit{Wrong\%} = fraction of incorrect claims (---\,=\,$n{<}3$).\\[4pt]
  \textsuperscript{c}~Contains unverifiable claims about callee behavior or project conventions.
\end{minipage}
\vspace{-10pt}
\end{table}


In this section, we motivate VERA through the poor performance of LLM-generated CoT vulnerability explanations, highlight the challenges of auditing LLM reasoning at scale, and describe how VERA addresses these challenges via SRR. 

LLM analyses may rely on pattern matching, memorization, or hallucinated reasoning~\cite{lekssays2025llmxcpg,shortcut2020,PearceA0DK22,turpin2023unfaithful,lyu2023faithful}. To evaluate reasoning validity, we manually audited free-form CoT explanations from four state-of-the-art LLMs on all 64 UAF vulnerable/patched function pairs in SVEN~\cite{SVEN} (see supplementary material for details).
Two human security experts audited each response across three dimensions: (1) verdict and CWE correctness, (2) factual accuracy of cited external knowledge (e.g., CVEs), and (3) the extent to which the reasoning was grounded in the code slice provided to the LLM as input. Our findings are three-fold, each illustrating a different way free-form CoT obscures reasoning mistakes that VERA is designed to surface.

First, despite achieving moderate binary accuracy ($46\text{--}60\%$), models rarely identify the correct vulnerability type (true CWE recall is only $3\text{--}16\%$) and label \emph{both} vulnerable and patched versions as \textsc{VULNERABLE} in up to $80\%$ of pairs. 
These numbers indicate widespread reasoning mistakes in the model's explanation, yet free-form CoT poses challenges for auditing. Natural language imposes no structure on claims, allows contradictions to coexist with correct statements, and cannot be reliably parsed or checked. VERA addresses this by replacing free-form prose with an SRR that explicitly records every claimed memory operation, state transition, and property check. This state-transition-based explanation mirrors conceptually how human experts reason about and explain vulnerabilities, by tracing where an object is allocated, freed, and accessed again. 
With reasoning expressed in this structured form, the multi-stage judge can automatically locate and check each claim against the code slice.

Second, examining the content of these errors more closely, we find that models frequently jump to conclusions by citing CVEs, bug reports, or domain conventions as justification rather than the code itself.
Our manual review showed these claims are \textit{factually wrong $55\text{--}61\%$ of the time}. For example, when analyzing the SVEN function \texttt{kvm\_ioctl\_create\_device}, Sonnet-5 flagged both the vulnerable and patched versions as \textsc{VULNERABLE} while citing CVE-2019-6974—a reference-counting defect, not a UAF. Free-form CoT permits authoritative assertions (e.g., \textit{``This matches CVE-X''}) without mapping them to concrete code regions, which may trick both human reviewers and LLM judges. 
VERA detects this by validating each claim recorded in the SRR against the input code. A verdict citing CVE-2019-6974 must also record the corresponding pointer allocation, deallocation, and dereference in the code slice. A hallucinated citation will produce a trace with mistakes, and VERA flags the reasoning mistakes.

Third, across all models, $19\text{--}34\%$ of responses relied on ungrounded assertions about callee behaviors or domain conventions, which is not part of the input to LLMs. For instance, when analyzing \texttt{disk\_seqf\_stop}, Sonnet-5 declared the slice \textsc{NOT VULNERABLE} by asserting that \textit{``the stop callback is always called with a valid seqf—this is an accepted kernel convention.''} In unstructured text, such premises are embedded into text, making it impossible to separate facts from speculative assumptions.
 VERA makes such assumptions explicit by encoding callee behavior and domain conventions as separate SRR fields, enabling judges to verify or reject them.
\section{Structured Vulnerability Reasoning}
\label{sec:structure}

In this section, we present the key components underlying VERA's reasoning audit framework. We first model vulnerabilities as sequences of state transitions within program slices (Sec.~\ref{sec:state_def}). This representation closely mirrors conceptually how human security experts reason about and explain vulnerabilities, and forms the foundation of the Structured Reasoning Record (SRR) schema (Sec.~\ref{sec:srr_schema}). The SRR is designed to externalize implicit LLM reasoning into a collection of inspectable facts. Finally, we introduce a failure taxonomy comprising eight fundamental reasoning error types, each corresponding to errors that may arise within SRR blocks (Sec.~\ref{sec:taxonomy}).

\subsection{State-based Vulnerability Definition}
\label{sec:state_def}

Following prior works that model program behavior as state transition systems and software vulnerabilities as sequences of state transitions that reach a bad state~\cite{Lamport1977, SLAM, Engler2000, Schneider2000, cheng2024fast,StromY86}, we specify software vulnerabilities using state-based definitions. 
For example, a use-after-free (UAF) vulnerability is characterized by a sequence of allocation, release, and access operations on a resource, where each operation transitions the resource to a new state. The vulnerability manifests when the resource is dereferenced after it has entered the \texttt{FREED} state~\cite{cheng2024fast}. The informal definitions for CWEs of concern are shown below.







\refstepcounter{definitionctr}
\label{def:mem}
\begin{tcolorbox}[
    float=htb,
  enhanced,
  colback=defbg,
  colframe=defframe,
  arc=4pt,
  boxrule=0.6pt,
  left=6pt, right=6pt, top=4pt, bottom=6pt,
  fontupper=\small,
  title={State-based Specification of C/C++ Memory Vulnerabilities},
  fonttitle=\small\bfseries\color{white},
  attach boxed title to top left={yshift=-2mm, xshift=6pt},
  boxed title style={
    colback=deftitle,
    colframe=deftitle,
    arc=3pt,
    boxrule=0pt,
    left=4pt, right=4pt,
  },
]

A memory resource $M$ exists in a state $S \in \{\texttt{NULL},\,\texttt{ALLOC},\,\texttt{FREED}\}$.
A pointer $p$ is a handle to $M$ and has a lexical scope. $M$ is \emph{reachable} iff some in-scope pointer holds its address. \limin{where is live defined?}

\smallskip
\noindent\textbf{Operations and State Transitions}:
\begin{itemize}[nosep, leftmargin=*, topsep=2pt]
  \item \textbf{Initial State}: the default state for $M$ is $\texttt{NULL}$.
  \item \textbf{Allocation} (\textit{malloc}, \textit{calloc}, \textit{realloc}, or a
        project-specific allocator): creates $M$ and sets $\textit{State}(M) = \texttt{ALLOC}$.
  \item \textbf{Deallocation} (\textit{free}, \textit{kfree}, or a project-specific release
        helper): if $p$ points to $M$, sets $\textit{State}(M) = \texttt{FREED}$.
        $\textit{free}(\texttt{NULL})$ is a defined no-op.
\end{itemize}

\smallskip
\noindent\textbf{Aliasing}: for any pointer $x$ s.t. $x$ points to $M$,
$\textit{State}(x) \equiv \textit{State}(M)$.

\smallskip
\noindent\textbf{Vulnerability (violation conditions)}:
\begin{itemize}[nosep, leftmargin=*, topsep=2pt]
  \item \textbf{CWE-416} (Use-After-Free): $\textit{Access}(x)$ --- a dereference, read, write, or
        pass --- when $\textit{State}(x) = \texttt{FREED}$.
  \item \textbf{CWE-415} (Double Free): $\textit{free}(x)$ when
        $\textit{State}(x) = \texttt{FREED}$.
  \item \textbf{CWE-401} (Memory Leak): $M$ becomes unreachable while $\textit{State}(M) = \texttt{ALLOC}$.
\end{itemize}

\end{tcolorbox}
This specification captures heap-memory lifecycle transitions.
A heap resource $M$ is accessed through one or more pointer handles $p$, where reachability depends on the existence of a handle in scope. 
State changes are induced by operations (e.g., $\mathtt{alloc}$ and $\mathtt{free}$). The aliasing rule ensures that any alternate pointer $x$ referencing $M$ 
inherits its state~\cite{FinkYDRG06}. 
Use-after-free accesses a resource in the \texttt{FREED} state. Double free attempts to release it a resource in the \texttt{FREED} state. A memory leak occurs when the resource becomes unreachable while the resource remains in \texttt{ALLOC} state.

This state-transition formulation generalizes naturally to stateful resource abstractions following open--use--close or state-gated patterns, such as network sockets, database connections, or session authentication states. 
Other natural extensions include, pointer validity, general resource releases, file descriptors, and object initialization (see definitions in supplementary material).  \limin{do we have supplementary material?}
Extending the framework to additional CWEs requires specifying their resource states, state-relevant operations, and violation conditions, alongside updating the reasoning schema and auditor rules. 

\subsection{Structured Reasoning Records}
\label{sec:srr_schema}

Standard free-text reasoning hides reasoning errors within fluent prose~\cite{lekssays2025llmxcpg,shortcut2020,PearceA0DK22,turpin2023unfaithful,lyu2023faithful}, rendering automated verification impossible. 
Based on state-transition-based vulnerability specifications, VERA requires the target detection model to structure its analysis into four linked blocks (Step~I of Fig.~\ref{fig:overview-vera}): scope and alias sets (Block~1), state traces along execution paths (Block~2), property evaluations (Block~3), and the final verdict (Block~4). Crucially, each block records the model's \textit{claims}. We demonstrate with the example in Fig.~\ref{fig:motivating}, the SRR is shown in Fig.~\ref{fig:record_srr}.

\definecolor{srrb1}{RGB}{237,244,255}   
\definecolor{srrb2}{RGB}{237,250,240}   
\definecolor{srrb3}{RGB}{255,248,232}   
\definecolor{srrb4}{RGB}{247,240,252}   

\newcommand{\srrsmall}{\fontsize{6.4}{7.6}\selectfont}
\newcommand{\srrkey}[1]{\texttt{\textbf{#1}}}

\tcbset{
  srrblk/.style={
    enhanced, sharp corners, boxrule=0.4pt,
    fonttitle=\bfseries\fontsize{6.4}{7.6}\selectfont, coltitle=black,
    top=2pt, bottom=2pt, left=3pt, right=3pt, toptitle=1pt, bottomtitle=1pt,
  },
}

\begin{figure}[t]
\centering
\srrsmall

\begin{minipage}[t]{0.30\linewidth}
\vspace{0pt}
\begin{tcolorbox}[srrblk, colback=srrb1, colframe=blue!45!black,
  colbacktitle=srrb1, title={Block~1 (Scope)}]
\srrkey{pointers}: \texttt{e} (\texttt{ALLOC})\\[2pt]
\srrkey{memory\_ops}:\\[1pt]
\begin{tabular}[t]{@{}c@{\ }c@{\ }c@{}}
\texttt{type} & \texttt{ptr} & \texttt{op}\\
\midrule
\texttt{alloc} & \texttt{e} & \texttt{cache\_lookup}\\
\red{\texttt{free}}   & \texttt{e} & \texttt{cache\_remove}\\
\texttt{deref}  & \texttt{e} & \texttt{e->refs}\\
\end{tabular}\\[2pt]
\red{Missing frees on L13-14, deref on L16}
\end{tcolorbox}
\end{minipage}\hfill
\begin{minipage}[t]{0.40\linewidth}
\vspace{0pt}
\begin{tcolorbox}[srrblk, colback=srrb2, colframe=green!40!black,
  colbacktitle=srrb2, title={Block~2 (Trace state changes)}]
\srrkey{paths}: \texttt{P0} (always);\;
P1 (\texttt{e == NULL});\;
P2 (\texttt{e != NULL});\; P3 (\texttt{e->refs ==0});\; P4 (\texttt{e->refs !=0})\\[2pt]
\srrkey{trace}:\\[1pt]
\begin{tabular}[t]{@{}c@{\ }c@{\ }c@{\ }c@{}}
\texttt{path} & \texttt{operation} & \texttt{state ch.}\\
\midrule
P0 & \texttt{alloc (cache\_lookup)} & e: \texttt{ALLOC}\\
\multirow{2}{*}{P2} & \texttt{\red{free} (cache\_remove)} & \red{e: \texttt{FREED}}\\
 & \texttt{deref (e->refs)} & e: \texttt{FREED}\\
P3, P4 & \multicolumn{2}{c}{Not executed, UAF}\\
\end{tabular}
\end{tcolorbox}
\end{minipage}\hfill
\begin{minipage}[t]{0.26\linewidth}
\vspace{0pt}
\begin{tcolorbox}[srrblk, colback=srrb3, colframe=orange!55!black,
  colbacktitle=srrb3, title={Block~3 (CWE property audit)}]
\srrkey{IDENTIFIED}\; CWE-416\\[2pt]
{\itshape remaining assessments omitted}
\end{tcolorbox}
\begin{tcolorbox}[srrblk, colback=srrb4, colframe=violet!55!black,
  colbacktitle=srrb4, title={Block~4 (Conclusion)}]
\srrkey{verdict} = \texttt{VULNERABLE}\\[2pt]
\srrkey{violation\_condition}: \texttt{CWE-416}\\[2pt]
{\itshape justification omitted}
\end{tcolorbox}
\end{minipage}

\vspace{-5pt}
\caption{Structured Reasoning Record (SRR) for the code snippet in 
Fig.~\ref{fig:motivating:code}. The four blocks organize the detection model's reasoning, including ungrounded assumptions and incorrect state transitions, into named, typed fields that VERA audits against the failure taxonomy. \red{Red} indicates errors detected by VERA.}
\label{fig:record_srr}
\end{figure}

\textit{Block~1 (Scope):} Captures declared entities, including relevant variables, assumed initial states, observed operations, and alias equivalence sets. In our running example (Fig.~\ref{fig:record_srr}), the model identifies pointer \texttt{e} with initial state \texttt{ALLOC} and lists operations \texttt{cache\_lookup}, \texttt{cache\_remove}, and \texttt{e->refs}, missing key free operations. This block can be viewed as the definitions of variables and operations in the explanation. 
\textit{Block~2 (Trace of State Changes):} Records the model's path-sensitive reasoning as an explicit sequence: which operations occur, along which execution paths, in what order, and with what state transitions. In Fig.~\ref{fig:record_srr}, the model asserts that $P_0$ executes unconditionally and branches at \texttt{if (!e)} into $P_1$ and $P_2$, then logs the resulting state changes. Because the model incorrectly assumes \texttt{cache\_remove} releases \texttt{e}, it records an incorrect transition from \texttt{ALLOC} to \texttt{FREED} on $P_1$ — a reasoning mistake now expressed as an explicit field that can be inspected and checked.
\textit{Block~3 (CWE Property Audit):} Records explicit per-CWE safety claims. 

In Fig.~\ref{fig:record_srr}, it asserts that CWE-416 (Use-After-Free) is \texttt{IDENTIFIED} as \texttt{e} was dereferenced in the \texttt{Freed} state, making the basis of its judgment inspectable.
\textit{Block~4 (Conclusion):} Records the final classification (\texttt{VULNERABLE} or \textsc{SAFE}) alongside the cited \texttt{violation\_condition}. By separating the conclusion from the reasoning in Blocks~1-3, VERA can check whether the verdict is coherent with the model's own intermediate claims — a consistency that free-form CoT never exposes.

By structuring intermediate reasoning into discrete, typed fields, the schema makes LLM vulnerability analyses auditable and surfaces errors as concrete, verifiable claims.

\begin{table*}[t]
\centering
\small
\setlength{\tabcolsep}{4pt}
\caption{Taxonomy of reasoning errors in state-based vulnerability analysis.}
\vspace{-5pt}
\label{tab:taxonomy}
\scalebox{0.73}{
\begin{tabular}{@{} l p{8cm} p{7cm} }
\toprule
\textbf{Name} & \textbf{Explanation} & \textbf{Example}  \\
\midrule
\multicolumn{3}{@{}l}{\textit{Lexical and evidence integrity}} \\[2pt]

$F_1$ Scope Error
  & Incorrect list of cwe-relevant variables and/or state-changing operations, and incorrect initial state of the variables.
  & Model overlooked \texttt{e} as a pointer.\\[6pt]

$F_2$ External Assumption
  & Unsupported claims about caller/callee behavior, or project conventions treated as an established fact.
  & Model infers that \texttt{cache\_remove} releases \texttt{e} based on its name or comments.\\[6pt]

\midrule
\multicolumn{3}{@{}l}{\textit{State correctness}} \\[2pt]

$F_3$ State Transition Error
  & Transition applied to the wrong variable or the wrong operation; the state change does not follow the operation; correct operations but incorrectly sequenced relative to source
  & Model records \texttt{free(e)} but notes the transition to \texttt{ALLOC} \\[6pt]

$F_4$ Alias Error
  & Missing or hallucinated alias, or states not correctly propagated to its co-referencing aliases.
  & Model marks \texttt{p1} as \textsc{freed} but leaves alias \texttt{p2} as \textsc{valid}, then concludes \texttt{p2} is safe to dereference.
  \\[6pt]

$F_5$Trace Incompleteness
  & Intermediate state changes are omitted between two trace steps, or a path dependency is referenced without being established in the preceding trace.
  & Model skips a \texttt{free(buf)} call between two steps tracking \texttt{buf}, presenting an incomplete view of its lifecycle.\\[6pt]

\midrule
\multicolumn{3}{@{}l}{\textit{Path Validity}} \\[2pt]

$F_6$ Path Fabrication
  & An execution path is asserted along a non-existent or unconnected branch, or incorrect activation conditions.
  & Model claims the error path executes when \texttt{ptr != NULL}, but the branch guard in source is \texttt{ptr == NULL} 
  \\[6pt]

$F_7$ Dead Code Reliance
  & A state transition is cited inside a statically unreachable block, e.g., a literal-false conditional, a preprocessor-disabled branch, or a provably unsatisfiable loop, treating dead code as live evidence.
  & Model uses a transition inside \texttt{if (0) \{ free(p); \}} as evidence that \texttt{p} is released on the error path.
 \\[6pt]

\midrule
\multicolumn{3}{@{}l}{\textit{Output coherence}} \\[2pt]

$F_8$ Verdict Inconsistency
  & The step-by-step property evaluation and the final verdict are in direct logical conflict.
  & Model marks CWE-416 as \textsc{violated} in Block~3 but concludes \texttt{NOT\_VULNERABLE} in Block~4, or marks all properties \textsc{satisfied} yet reports \texttt{VULNERABLE}.
 \\[4pt]

\bottomrule
\end{tabular}}
\vspace{-10pt}
\end{table*}

\subsection{Reasoning Failure Taxonomy}
\label{sec:taxonomy}
To systematically identify LLM reasoning errors as presented in SRR, we derive a taxonomy of eight fundamental reasoning failure modes ($F_1$--$F_8$), see Tbl.~\ref{tab:taxonomy}. These failure modes 
correspond to the state-based definition and the SRR schema: the tracked resources and active handles, the valid state transitions, the paths containing those transitions, and the violation conditions.
Hence, our taxonomy is organized across four core reasoning dimensions:
\textit{1)~Lexical and evidence integrity} ($F_1, F_2$), whether all relevant variables, pointers, and initial states are correctly scoped and that every claimed fact is anchored to concrete source operations rather than external assumptions or identifier naming conventions;
\textit{2)~State correctness} ($F_3, F_4, F_5$), whether state transitions accurately match witnessed source operations, that alias relations are tracked and state mutations propagated across all co-referencing handles~\cite{FinkYDRG06}, and that intermediate state changes are completely recorded without omissions;
\textit{3)~Path validity} ($F_6, F_7$), whether state transitions occur along valid, connected execution paths, requiring these paths to be syntactically connected from entry to the transition point, with branch conditions faithfully drawn from source rather than fabricated branches or statically dead code; and
\textit{4)~Output coherence} ($F_8$), whether the final security verdict strictly follows from step-by-step property evaluations without logical self-contradictions~
\cite{turpin2023unfaithful}.

Failure modes $F_1, F_2, F_3, F_5, F_6$, and $F_7$ check for \textit{code adherence}: every symbol, transition, path, and omission is faithfully grounded in the code slice. $F_4$ and  $F_8$ check for \textit{internal consistency}: alias propagation is coherent and the final verdict strictly aligns with property evaluations.
All these failures can be detected by inspecting the SRR. Checking Block~1--2 against the source code identify lexical and evidence integrity failures ($F_1$--$F_2$) and alias identification ($F_4$); state correctness ($F_3$--$F_5$) and path validity ($F_6$--$F_7$) can be detected through Block~2; and verdict coherence ($F_8$) is checked by comparing the verdict against the reasoning trace in Blocks~2--4.

\section{Auditing LLM Vulnerability Reasoning}
\label{sec:judge}
After the detection model outputs an SRR, VERA evaluates it using the multi-stage judge shown in Step~II of Fig.~\ref{fig:overview-vera}. 
The judge performs a 
\textit{claim validation}: it checks whether the model's explicit reasoning claims are grounded in the input code and internally coherent. 
%
The multi-stage judge contains eight stages (S1--S8), each targeting one failure mode ($F_1$--$F_8$ from Tbl.~\ref{tab:taxonomy}). 
The stages follow a logical dependency chain: S1--S2 establish scope; S3--S4 validate path structure against the declared scope; S5--S6 check state transitions against the established scope and paths; S7 verifies alias propagation across the full trace; and S8 checks verdict coherence against all prior findings.

\boldparagraph{S1 (deterministic): Checking for Scope Error ($F_1$)}
Stage~S1 checks Block~1 against the source code AST to verify the completeness and accuracy of the declared scope.
Using Tree-Sitter, VERA extracts two ground-truth sets: cwe-relevant variables (i.e., memory-related variables such as pointers and fields for memory CWEs, and resource-related variables such as file descriptor for resource CWEs) and state-relevant operations (e.g., allocations, deallocations, dereferences for memory CWEs, acquire and release for resource CWEs). First, we check for scope hallucination and completeness, every variable and operation claimed in Block~1 exist in the input code slice, and all variables and operations in the code slices must exist in the declared scope in Block~1. Operations hidden inside missing callee context are flagged with a \texttt{MISSING\_CONTEXT} warning. 
In Block~1 of Fig.~\ref{fig:record_srr}, S1 would flags the missing deallocation operations on L13--14 and the missing dereference on L16 as scope incompleteness errors. 
Additionally, initial state declarations in Block~1 must comply with C memory initialization rules: claiming \texttt{ALLOCATED} or \texttt{FREED} for a local variable before any corresponding operation appears in the slice is flagged as a scope error. Function parameters are an exception: when caller context is absent from the slice, their initial states cannot be checked hence not flagged as a scope error but a \texttt{MISSING\_CONTEXT} warning.

\boldparagraph{S2 (LLM): Check for Assumptions ($F_2$)}
Stage~S2 takes Block~1 and Block~2 together with the input code slice and uses targeted LLM calls to check whether any claims rely on hidden or external assumptions ungrounded in the code. 
For each claim, a targeted LLM call evaluates whether the provided justification is backed strictly by evidence explicitly present in the reasoning record and source code slice. In Fig.~\ref{fig:record_srr}, S2 will flag the claim that \texttt{cache\_remove} releases \texttt{e} as an ungrounded external assumption since no deallocation operation on \texttt{e} exist in the code slice. This step deploys LLM because detecting implicit natural-language assumptions in justifications is an irreducibly semantic task that cannot be resolved via deterministic parsing. \ronghao{I guess readers will have confusions on whether cache\_remove really frees the pointer or not? If not, this is an LLM reasoning error. If yes, then LLM has no way to output correct reasonings?} \caroline{cache\_remove is not freeing e, added more obvious annotation in the figure}

\boldparagraph{S3 (deterministic): Check for Path Identification ($F_6$)}
Stage~S3 checks Block~2's trace against Block~2's declared paths to verify that every referenced Path ID is explicitly declared. 
State transitions are only valid along realizable execution paths.  Referencing an undeclared Path ID indicates a structural inconsistency and is flagged as an $F_6$ path fabrication.

\boldparagraph{S4 (deterministic): Check for Dead Code Reliance ($F_7$)}
Stage~S4 checks Block~2's trace against the source AST to detect whether any claimed state transitions are grounded in syntactically unexecutable code.
S4 targets syntactically detectable dead code (e.g., \texttt{if (0)}, \texttt{while (false)}) — the subset reliably identifiable with static analysis on a code slice alone. More accurate dead-code detection would require CFG construction and is left for future work.

\boldparagraph{S5 (deterministic): Check for Trace Coverage ($F_5$)}
Stage~S5 cross-checks Block~2's execution trace against the declared operations in Block~1 to enforce internal consistency. While S1 grounds Block~1 against the source code, S5 ensures the model does not cherry-pick supporting operations in its trace while silently omitting contradicting ones. To avoid false positives, VERA restricts this check to variables actively tracked in the trace: S5 flags a gap only when a model traces a variable but omits a confirmed operation on that same variable. For instance, if the model traces pointer \texttt{p} through allocation and dereference but skips an intermediate \texttt{free(p)} call, S5 flags the omission—as that missing operation determines whether the pattern constitutes a Use-After-Free. As illustrated in Fig.~\ref{fig:record_srr}, S5 does not re-flag ungrounded operations already caught by S1, focusing strictly on trace-completeness gaps.


\boldparagraph{S6 (deterministic): Check for State Transition ($F_3$)}
Stage~S6 checks every state transition in Block~2 against the source code and CWE state-based definition to verify correctness and ordering. 
For every state transition in Block~2, Stage~S6 deterministically checks whether the state update complies with the transitions defined in the state-based definitions, e.g., transitioning to \texttt{FREED} on a free operation; and whether the ordering of transitions is consistent with the input code slice. Additionally, for resource leaks (CWE-401 and CWE-772), an end-of-trace check verifies whether allocated variables remain in an unreleased state at the end of the function. In Fig.~\ref{fig:record_srr}, S6 flags the \texttt{ALLOC} to \texttt{FREED} transition attributed to \texttt{cache\_remove}, since no deallocation semantics are established for that call.

\boldparagraph{S7 (LLM): Check for Alias ($F_4$)}
Stage~S7 takes Block~1 and Block~2 together with the source slice and uses a targeted LLM call to evaluate whether alias relationships are correctly identified and state mutations propagated to all coreferencing aliases. 
For example, if \texttt{p} is freed and \texttt{q = p}, then both \texttt{p} and \texttt{q} should be in the \texttt{FREED} state. Because aliasing can arise through patterns that require whole-program points-to analysis to resolve precisely, including pointer arithmetic, struct field access, type punning, or values returned from opaque callees, S7 deploys a targeted LLM call rather than deterministic AST matching.

\boldparagraph{S8 (deterministic+LLM): Check for Verdict Consistency ($F_8$)}
Stage~S8 checks Blocks~2--4 to validate coherence between the reasoning trace, the CWE 
claims, and the final verdict. 
First, we check whether each claim is backed by the correct state transitions (e.g., CWE-416 requiring a dereference step following a \texttt{FREED} state). Safe functions contain no violation claims; for these, trace correctness is checked in prior stages S3 to S6. Second, we confirm the natural-language rationale in Block~3 for final verdict aligns with the state transition trace in Block~2, via a targeted LLM call. Third, we check deterministically that Block~4's binary verdict is logically consistent with Block~3's CWE violation claims, i.e., vulnerable if any CWE identified, safe otherwise. 

\boldparagraph{Final Judge Outcome} 
Upon executing all eight stages, VERA classifies the reasoning record into one of four final outcomes.
(1)~\textbf{\texttt{NO\_ERROR}}: All stages pass without hard failures, indicating the reasoning chain is complete, internally consistent, and grounded in the input code slice.
(2)~\textbf{\texttt{MINOR\_ERRORS}}: At least one stage triggers a warning without hard failures (e.g., unverifiable callee context or unconfirmed aliases). The core reasoning chain remains valid, but requires targeted review of the claims that triggered the warning.
(3)~\textbf{\texttt{MAJOR\_ERRORS}}: At least one stage triggers a hard failure ($F_1$--$F_8$), such as a claimed dereference absent from the code slice ($F_3$), an inverted allocation--free sequence ($F_4$), or a verdict inconsistent with property checks ($F_7$). VERA reports the specific stage and offending claim involved. Human security analysts can use these localized diagnostics to quickly audit suspicious code regions without reading full traces, while downstream LLMs can ingest these targeted error reports to drive self-correction loops or filter unreliable verdicts. Crucially, a correct binary verdict and \texttt{MAJOR\_ERRORS} can co-occur, exposing instances where a model reached the right answer through fabricated reasoning.
(4)~\textbf{\texttt{INCOMPLETE\_RECORD}}: The model failed to generate a parseable reasoning record, excluded from reasoning-quality evaluation and flagged for re-query.


\boldparagraph{Evaluating the Judge via Mutation Testing}
Assessing the reliability of reasoning judges 
requires ground-truth reasoning errors~\cite{zhu2026vulinstruct}, which is hard to obtain. Human review of individual reasoning chains is expensive and does not scale. 
One approach to acquire ground-truth data at scale in software testing without human annotation is mutation testing, which generates mutants with errors injected~\cite{doddapaneni2024fbi,delcourt2026breaking}. While this approach does not transfer directly to natural language reasoning,
 VERA's SRR can be programmatically injected with precise, category-controlled faults. For example, deleting a state transition ($F_3$), severing an alias link ($F_4$), or fabricating an execution path ($F_6$) can produce exact ground-truth errors. 
 VERA thus supports automated mutation-based evaluation of judges across the full failure taxonomy ($F_1$--$F_8$), enabling rigorous, scalable stress testing of reasoning judges operating on SRRs.

\boldparagraph{Scope and Limitations}
To enable language-agnostic, build-free deployment at scale, VERA's judge operates directly on uncompiled code slices without requiring whole-program build environments. This design choice imposes two main operational boundaries. First, implicit operations introduced through type casts, struct field accesses, or opaque callees require type-system reasoning beyond AST-level pattern matching. Consequently, the completeness check in S1 is scoped to statically verifiable operations — explicit calls to known C/C++ memory functions (e.g., \texttt{malloc}, \texttt{free}, \texttt{realloc}); implicit operations are conservatively treated as warnings rather than hard failures. This is a deliberate trade-off that preserves zero false positives on structural checks at the cost of moderate recall in Stages~S1 and S5. Second, checks that fundamentally require whole-program compilation, such as precise dead-code detection, alias resolution beyond direct pointer assignment, and formal path reachability, are currently outside VERA's current scope. These are fundamental boundaries of intra-procedural auditing; integration with whole-program control-flow and points-to solvers is a natural direction for future work.

Together, the eight-stage pipeline and mutation testing form a self-contained auditing framework: the judge measures reasoning quality independently of verdict correctness, and mutation testing evaluates the judge itself without requiring ground truth.

\section{Evaluation}
\label{sec:evaluation}
 
We evaluate VERA and answer the following questions. 
To establish that our error taxonomy is empirically grounded, we check that the failure categories reflect errors that occur in real LLM reasoning and answer: (\textbf{RQ1}) Do the failure categories in our taxonomy reflect real LLM reasoning failures? 
Next, with the taxonomy established, we evaluate the effectiveness of VERA's judge in detecting these errors. 
Since no ground-truth labels exist for reasoning errors, we use mutation testing to inject known errors into valid reasoning records, enabling precise, controlled evaluation of judge effectiveness. We answer 
(\textbf{RQ2}) How does VERA's multi-stage hybrid judge compare to a stage-structured LLM-as-a-judge in detecting reasoning errors? Finally, having established that VERA reliably detects reasoning errors, we then apply it at scale to real LLM reasoning traces to identify systematic reasoning failures, including ``correct verdict, flawed reasoning''. We compare against free-form CoT baselines and answer: (RQ3) What systematic reasoning failures does VERA expose that binary accuracy metrics and free-form CoT judges fail to detect?"


\subsection{Experimental Setup}
\boldparagraph{Datasets} We evaluate VERA across diverse real-world vulnerability benchmarks SVEN~\cite{SVEN} and Primevul~\cite{Ding2024PrimeVul} containing paired vulnerable and fixed function versions across five CWE categories in C/C++: CWE-416 Use-After-Free ($184$ pairs: $64$ SVEN, $100$ PrimeVul, $20$ newly collected), CWE-476 NULL Pointer Dereference ($277$ pairs: $77$ SVEN, $100$ PrimeVul, $100$ newly collected), CWE-401 Memory Leak ($200$ pairs: $100$ PrimeVul, $100$ newly collected), CWE-415 Double Free ($110$ pairs: $65$ PrimeVul, $45$ newly collected), and CWE-772 Unreleased Resource ($80$ pairs: $76$ PrimeVul, $4$ newly collected). We include paired datasets to evaluate whether stage-level reasoning failures persist across a vulnerable function and its patched counterpart, motivated by our observation in Sec.~\ref{sec:motivation}, exposing whether model errors are driven by code structure rather than vulnerability presence.

\boldparagraph{Models} Our experimental pipeline involves two roles for LLMs: detection models that generate vulnerability reasoning and judge models that evaluate reasoning correctness. For vulnerability detection, unlike in Sec.~\ref{sec:motivation}, we used SotA models to establish that the problem exists even with the most frontier models, here we evaluates across a broader, varied set of open and closed-source models to assess generalizability. We included results from six state-of-the-art LLMs: sonnet-5~\cite{anthropic2026claude}, gemma-4-26b~\cite{google2026gemma}, llama-4-maverick~\cite{meta2026llama}, nemotron-3-nano~\cite{nvidia2026nemotron}, glm-5.3-flash~\cite{zhipu2026glm}, qwen-3.8-27b~\cite{qwen2026qwen}, spanning open-source and proprietary models with different sizes\footnote{Due to page limit, we present a subset of the result in the paper, the rest can be found in supplementary}. For reasoning evaluation, we employ three frontier and cost-efficient judge models from different providers than detection models: GPT-4.1 (OpenAI)~\cite{openai2024gpt4technicalreport}, Mistral-Large-2512 (Mistral AI)~\cite{mistral2025large}, and DeepSeek-V3 (DeepSeek)~\cite{deepseek2025chat}. 

\boldparagraph{Evaluation Configurations} 
Because no prior vulnerability reasoning auditing system directly audits intermediate reasoning without relying solely on an LLM-as-judge, we establish four evaluation configurations to systematically isolate the contributions of two core design choices: \textit{reasoning format} (free-Form CoT~\cite{Wei2022ChainOfThought} vs.\ structured SRR) and \textit{audit architecture} (holistic LLM-as-judge~\cite{Zheng2023JudgingLLM} vs.\ stage-structured LLM vs.\ VERA's multi-stage hybrid judge). 
1) Free-form baseline \textbf{B1}: \textit{Free-Form CoT + Holistic LLM Judge.}
The detection model is prompted for open-ended reasoning. A holistic LLM-as-a-judge inspects the full narrative without stage decomposition.
2) Free-form baseline \textbf{B2}: \textit{Free-Form CoT + stage-structured LLM Judge.}
Same free-form detection output as B1, but the judge is prompted to evaluate reasoning explicitly with stages corresponding to the eight taxonomy categories.
3) LLM-as-judge baseline \textbf{B3}: \textit{SRR + stage-structured LLM Judge.}
The detection model materializes its reasoning into VERA's SRR, then audited purely via an LLM judge using the same stage-structured prompts as B2.
4) \textbf{VERA}: \textit{SRR + Multi-Stage Hybrid Judge.} 
The detection model outputs an SRR schema, which is then audited by VERA's multi-stage hybrid pipeline.
We show the benefit of structured reasoning by comparing \textbf{B1/B2 vs.\ VERA} (RQ3), and show the effectiveness of our multi-stage judge by comparing \textbf{B3 vs.\ VERA} (RQ2).
\begin{table}[t]
\centering
\caption{Mapping between reasoning failure types and each mutation operator.}
\vspace{-5pt}
\label{tab:mutation_mapping}
\small
\scalebox{0.9}{
\begin{tabular}{ l p{11cm}}
\toprule
 \textbf{Reasoning Failure} & \textbf{Mutation Operators} \\
\midrule
$F_1$ scope error
  & $M_{1a}$: add a nonexistent variable or remove a random variable; $M_{1b}$: add a nonexistent operation or remove a random operation;
    $M_{1c}$: introduce incorrect initial state \\
$F_2$ ungrounded assumptions
  & $M_2$: inject ungrounded external assumption into reasoning prose \\
$F_3$ incorrect state transitions
  & $M_3$: invert state-transition ordering or, insert spurious free step on leaked variable (included for CWE-401) \\
$F_4$ incorrect aliases
  & $M_4$: add or erase the trace steps identifying aliases\\
$F_5$ incomplete trace
  & $M_{5a}$: make trace step references undeclared path ID;
    $M_{5b}$: add or drop one random trace step\\
$F_6$ fabricated path
  & $M_{6a}$: intentionally make path-activation condition vague;
    $M_{6b}$: make trace step reference undeclared path ID;
    $M_{6c}$: make path activation reference undeclared path ID \\
$F_7$ unreachable code
  & $M_7$: wrap trace evidence in syntactically dead block \\
$F_8$ inconsistent verdict
  & $M_{8a}$: flip verdict;
    $M_{8b}$: flip CWE status and verdict \\
\bottomrule
\end{tabular}
}

\end{table}

\boldparagraph{Mutation-based Judge Evaluation}
We designed 13 mutation operators mapped directly across our taxonomy stages (see Tbl.~\ref{tab:mutation_mapping}). Each operator targets the precise reasoning breakdown that a given stage is designed to catch. These operators are designed to simulate errors that naturally arise in LLM-generated structured reasoning records, e.g., variables that appear in the analysis but not in source, memory operations that are silently omitted, paths with vague or self-contradictory activation conditions, and alias relationships left untracked. We aim for completeness within each stage by covering both directions of error where applicable — for instance, hallucination (claiming absent elements) and omission (missing real ones) at S1, or missing trace steps and invalid path references at S5. Then, we measure detection performance with two metrics. Note that not all operators are applicable to all SRRs, e.g., no op identified $M_2$ cannot be applied. \textbf{Recall} is the fraction of mutants targeting a specific stage that are correctly flagged as failures. \textbf{False-positive rate} (FP) is the fraction of mutants targeting a \emph{different} stage that are incorrectly identified.

\definecolor{judgeShade}{gray}{0.95}
\begin{table*}[t]
\caption{RQ1: error labels obtained with open-ended error enumeration mapped to our error taxonomy and mutation operators. Error label occurrence counts obtained with three judge models across 3000 reasoning records produced by six detection models. A record may receive multiple labels. The bottom section lists the 21 instances ($<$\,0.05\%) not assignable to any class; the eight categories together cover $>$\,99.95\% of all observed error instances.}
\vspace{-5pt}
\label{tab:rq1}
\centering
\setlength{\tabcolsep}{3pt}
\resizebox{\linewidth}{!}{%
\begin{tabular}{llrrrrr@{\hspace{4pt}}rrrrr@{\hspace{4pt}}rrrrr}
\toprule
 && \multicolumn{5}{c}{\textbf{GPT}}
 & \multicolumn{5}{c}{\textbf{Mistral}}
 & \multicolumn{5}{c}{\textbf{DeepSeek}} \\
\cmidrule(lr){3-7}\cmidrule(lr){8-12}\cmidrule(lr){13-17}
\textbf{Failure category / Mutation operator} & \textbf{Representative error labels} &
  \textbf{416} & \textbf{476} & \textbf{415} & \textbf{401} & \textbf{772} &
  \textbf{416} & \textbf{476} & \textbf{415} & \textbf{401} & \textbf{772} &
  \textbf{416} & \textbf{476} & \textbf{415} & \textbf{401} & \textbf{772} \\
\midrule
%
\rowcolor{judgeShade}
\textbf{F1~~Scope Error} &
  & \textbf{565} & \textbf{807} & \textbf{517} & \textbf{558} & \textbf{328} & \textbf{365} & \textbf{508} & \textbf{295} & \textbf{338} & \textbf{182} & \textbf{1056} & \textbf{1465} & \textbf{1007} & \textbf{1074} & \textbf{580} \\
\quad $M_{1a}$\hphantom{b}~~Add/remove identifier & \texttt{phantom/missing\_variable}
  &  479 & 678 & 454 & 433 & 254 & 311 & 454 & 241 & 272 & 149 & 675 & 949 & 639 & 625 & 324 \\
\quad $M_{1b}$\hphantom{b}~~Add/remove op & \texttt{phantom\_op},\texttt{missing\_memory\_op}
  & 85 & 125 & 62 & 123 & 74 & 54 & 54 & 54 & 66 & 33 & 369 & 481 & 353 & 433 & 250 \\
\quad $M_{1c}$~~Wrong initial state & \texttt{incorrect\_initial\_state}
  &   1 & 4 & 1 & 2 & 0 & 0 & 0 & 0 & 0 & 0 & 12 & 35 & 15 & 16 & 6 \\
\midrule
%
\rowcolor{judgeShade}
\textbf{F2~~External Assumption} &
  & \textbf{77} & \textbf{94} & \textbf{84} & \textbf{100} & \textbf{66} & \textbf{202} & \textbf{291} & \textbf{149} & \textbf{166} & \textbf{66} & \textbf{627} & \textbf{878} & \textbf{583} & \textbf{586} & \textbf{298} \\
\quad $M_2$\hphantom{b}~~Ungrounded assumption& \texttt{unsupported\_claim}
  & 77 & 94 & 84 & 100 & 66 & 202 & 291 & 149 & 166 & 66 & 627 & 878 & 583 & 586 & 298 \\
\midrule
%
\rowcolor{judgeShade}
\textbf{F3~~State Hallucination} &
  & \textbf{363} & \textbf{381} & \textbf{352} & \textbf{289} & \textbf{152} & \textbf{276} & \textbf{323} & \textbf{293} & \textbf{272} & \textbf{114} & \textbf{301} & \textbf{297} & \textbf{283} & \textbf{233} & \textbf{94} \\
\quad $M_3$ \hphantom{b}~~Break state-transition order& \texttt{wrong\_op\_order}
  &363 & 381 & 352 & 289 & 152 & 276 & 323 & 293 & 272 & 114 & 301 & 297 & 283 & 233 & 94 \\
\midrule
%
\rowcolor{judgeShade}
\textbf{F4~~Alias Incompleteness} &
  & \textbf{39} & \textbf{45} & \textbf{29} & \textbf{20} & \textbf{8} & \textbf{24} & \textbf{21} & \textbf{22} & \textbf{31} & \textbf{10} & \textbf{315} & \textbf{260} & \textbf{207} & \textbf{206} & \textbf{103} \\
\quad $M_4$\hphantom{b}~~Add/Remove alias traces& \texttt{alias\_confusion}
 &  39 & 45 & 29 & 20 & 8 & 24 & 21 & 22 & 31 & 10 & 315 & 260 & 207 & 206 & 103 \\
\midrule
%
\rowcolor{judgeShade}
\textbf{F5~~Trace Incompleteness} &
  & \textbf{636} & \textbf{910} & \textbf{588} & \textbf{596} & \textbf{296} & \textbf{633} & \textbf{888} & \textbf{592} & \textbf{590} & \textbf{301} & \textbf{881} & \textbf{1260} & \textbf{786} & \textbf{763} & \textbf{398} \\
\quad $M_5$\hphantom{b}~~Add/Drop trace step for op& \texttt{missing\_trace\_step}
  & 636 & 910 & 588 & 596 & 296 & 633 & 888 & 592 & 590 & 301 & 881 & 1260 & 786 & 763 & 398 \\
\midrule
%
\rowcolor{judgeShade}
\textbf{F6~~Path Fabrication} &
  & \textbf{430} & \textbf{704} & \textbf{369} & \textbf{351} & \textbf{170} & \textbf{611} & \textbf{866} & \textbf{567} & \textbf{565} & \textbf{288} & \textbf{800} & \textbf{1155} & \textbf{737} & \textbf{767} & \textbf{407} \\
\quad $M_{6a}$\hphantom{b}~~Vague path-act. condition& \texttt{vague\_path\_cond}
  &421 & 695 & 364 & 340 & 168 & 611 & 865 & 566 & 565 & 288 & 641 & 918 & 604 & 610 & 313 \\
\quad $M_{6b}$\hphantom{b}~~Trace step ref. undeclared path& \texttt{path\_id\_mismatch}
  &  9 & 9 & 5 & 11 & 2 & 0 & 1 & 1 & 0 & 0 & 158 & 234 & 133 & 157 & 94 \\
\quad $M_{6c}$~~Path ref. undeclared path id & \texttt{phantom\_path}
  &   0 & 0 & 0 & 0 & 0 & 0 & 0 & 0 & 0 & 0 & 1 & 3 & 0 & 0 & 0 \\
\midrule
%
\rowcolor{judgeShade}
\textbf{F7~~Dead Code Reliance} &
  & \textbf{2} & \textbf{9} & \textbf{5} & \textbf{3} & \textbf{2} & \textbf{6} & \textbf{8} & \textbf{8} & \textbf{4} & \textbf{2} & \textbf{70} & \textbf{84} & \textbf{45} & \textbf{65} & \textbf{72} \\

\quad $M_{7}$\hphantom{b}~~Wrap trace evidence in dead code & \texttt{dead\_code\_executed}
  &  2 & 9 & 5 & 3 & 2 & 6 & 8 & 8 & 4 & 2 & 70 & 84 & 45 & 65 & 72 \\
\midrule
%
\rowcolor{judgeShade}
\textbf{F8~~Verdict Inconsistency} &
  & \textbf{13} & \textbf{30} & \textbf{20} & \textbf{40} & \textbf{49} & \textbf{71} & \textbf{99} & \textbf{90} & \textbf{83} & \textbf{79} & \textbf{444} & \textbf{676} & \textbf{392} & \textbf{434} & \textbf{259} \\
\quad $M_{8a}$~~Flip verdict & \texttt{verdict\_contradicts\_audit}
  & 13 & 29 & 18 & 38 & 45 & 71 & 99 & 90 & 83 & 79 & 437 & 649 & 389 & 426 & 251 \\
\quad $M_{8a}$~~Flip CWE violation and verdict & \texttt{incons.\ property\_audit}
  &  0 & 1 & 2 & 2 & 4 & 0 & 0 & 0 & 0 & 0 & 7 & 27 & 3 & 8 & 8 \\
\midrule
\rowcolor{judgeShade}
\textbf{Total mapped}&
  & \textbf{2125} & \textbf{2980} & \textbf{1964} & \textbf{1957} & \textbf{1071} & \textbf{2188} & \textbf{3004} & \textbf{2016} & \textbf{2049} & \textbf{1042} & \textbf{4494} & \textbf{6075} & \textbf{4040} & \textbf{4128} & \textbf{2211} \\
\midrule
\multicolumn{16}{l}{\textit{Unmapped labels (21~instances, $<$\,0.05\%)}} \\
\textit{Format, not reasoning}
  & \texttt{wrong\_step\_numbering} & 0 & 0 & 0 & 0 & 0 & 0 & 0 & 0 & 0 & 0 & 1 & 7 & 1 & 1 & 1 \\
\textit{One-time instance-specific}
  & \texttt{wrong\_op\_type\_mutex\_unlock} & 0 & 0 & 2 & 0 & 0 & 0 & 0 & 0 & 0 & 0 & 0 & 1 & 3 & 3 & 1 \\
\bottomrule
\end{tabular}}

\end{table*}

\subsection{RQ1: Taxonomy Coverage} 
To answer \textbf{RQ1}, we evaluate whether our taxonomy ($F_1\text{--}F_8$) accurately captures naturally occurring reasoning errors and whether our mutation operators ($M_1\text{--}M_{8}$) provide complete coverage over real-world failure modes. We sampled 50 natural, un-injected SRRs per CWE per dataset (10 total with $\ge 50$ functions: SVEN-416/476; PrimeVul-416/476/401/415/772; new-415/476/401) produced by each of the six detection models (total $50\times 10 \times 6 = 3000$ records sampled). To discover reasoning failures without biasing the models toward VERA, we prompted our three judge models with an open-ended error enumeration prompt, instructing them to describe flaws, hallucinations, or logical inconsistencies without any prior. 

Across all three judges, open-ended annotation yielded 324 distinct error labels with 41,365 total occurrences. As detailed in Tbl.~\ref{tab:rq1}, 313 of the raw labels mapped cleanly onto our taxonomy classes ($F_1\text{--}F_8$) and mutation operators ($M_1\text{--}M_{8}$). The vast majority mapped via direct synonymy—for instance, labels describing missing entities (\emph{e.g.}, \texttt{missing\_memory\_op}) or phantom variables (\emph{e.g.}, \texttt{phantom\_variable}) mapped to $F_1$ (Scope/Operations), while \texttt{omitted\_guard} mapped to $F_5$ (Trace Incompleteness). While each $F_i$ defines a failure category, mutation operators instantiate the concrete mechanisms by which that failure arises, for example, $F_1$ (scope error) can manifest as a missing variable, a hallucinated variable, or a wrong pointer target, each should be covered by a distinct operator. All labels mapped to $F_1$--$F_8$ are fully covered by our 13 mutation operators without requiring additional mutation operators. While this does not imply an exhaustive enumeration of all possible code mutations, it demonstrates that $M_1\text{--}M_{8}$ comprehensively represent all different types of naturally occurring reasoning errors required to evaluate judge effectiveness.
The remaining 21 instances ($\le0.05\%$ covering 11 distinct labels) fall into two categories. Out of these, 11 are output-format mistakes rather than reasoning errors (e.g., misnumbered trace steps). The other 10 are single-occurrence, function-specific labels (e.g., \texttt{incorrect\_mutex} for one particular lock pattern) that do not recur across functions or CWEs, and 8/10 are produced by a single model DeepSeek. These are consistent with label noise introduced by open-ended enumeration prompts. Reasoning error \emph{classes} require recurring, generalizable failure patterns; one-time observations about specific implementation details do not qualify. Across 3,000 records and three judge models, no recurring pattern emerged that our eight-class taxonomy does not cover. Additionally, judge models disagreed on the frequency of individual error types, Mistral flagged substantially more $F_2$, $F_4$, and $F_8$ instances than GPT-4.1 or DeepSeek, further motivating VERA's deterministic stage checks over pure LLM-as-judge evaluation.


Hence our results confirm that our failure taxonomy and mutation suite are both empirically grounded and comprehensive, answering RQ1.

\begin{table*}[t]
\centering
\caption{Per-stage mutation detection (recall\,/\,FP) across five CWE classes, data from different datasets aggregated.
  VERA vs.\ B3 (SRR + structured LLM judge).
  Judge models: GPT\,=\,GPT-4.1, DS\,=\,DeepSeek, Mis\,=\,Mistral. $N$: number of mutants; -: no mutant generated.
  \textbf{Bold}: recall\,=\,1.00 or FP\,=\,0.00.
  \colorbox{gray!15}{Gray cells}: LLM-dependent (all B3 cells;
  VERA S2, S7 which are primarily LLM). }
  \vspace{-5pt}

\label{tab:rq2}
\setlength{\tabcolsep}{3pt}
\scalebox{0.62}{
\begin{tabular}{l @{\,} l | ccc | ccc | ccc | ccc | ccc}
\toprule
\textbf{Stage} & & \multicolumn{3}{c}{\textbf{CWE-416} ($N\!=\!711$)} & \multicolumn{3}{c}{\textbf{CWE-476} ($N\!=\!1006$)} & \multicolumn{3}{c}{\textbf{CWE-415} ($N\!=\!332$)} & \multicolumn{3}{c}{\textbf{CWE-401} ($N\!=\!541$)} & \multicolumn{3}{c}{\textbf{CWE-772} ($N\!=\!178$)} \\
\cmidrule(lr){3-5}\cmidrule(lr){6-8}\cmidrule(lr){9-11}\cmidrule(lr){12-14}\cmidrule(lr){15-17}
 & & \textit{GPT} & \textit{DS} & \textit{Mis} & \textit{GPT} & \textit{DS} & \textit{Mis} & \textit{GPT} & \textit{DS} & \textit{Mis} & \textit{GPT} & \textit{DS} & \textit{Mis} & \textit{GPT} & \textit{DS} & \textit{Mis} \\
\midrule
\textit{S1 Scope} & VERA & \multicolumn{3}{c|}{.53/{\bf .00}} & \multicolumn{3}{c|}{.56/{\bf .00}} & \multicolumn{3}{c|}{.68/{\bf .00}} & \multicolumn{3}{c|}{.61/{\bf .00}} & \multicolumn{3}{c}{.68/{\bf .00}} \\
 & \cellcolor{gray!15}B3 & \cellcolor{gray!15}.62/.13 & \cellcolor{gray!15}.69/.30 & \cellcolor{gray!15}.98/.01 & \cellcolor{gray!15}.64/.04 & \cellcolor{gray!15}.77/.19 & \cellcolor{gray!15}{\bf 1.00}/{\bf .00} & \cellcolor{gray!15}.53/.15 & \cellcolor{gray!15}.73/.20 & \cellcolor{gray!15}{\bf 1.00}/{\bf .00} & \cellcolor{gray!15}.75/.11 & \cellcolor{gray!15}.74/.24 & \cellcolor{gray!15}{\bf 1.00}/{\bf .00} & \cellcolor{gray!15}.63/.03 & \cellcolor{gray!15}.79/.17 & \cellcolor{gray!15}{\bf 1.00}/.02 \\
\addlinespace[1pt]
\hline
\textit{S2 Evidence} & VERA & \cellcolor{gray!15}.33/.06 & \cellcolor{gray!15}.41/.11 & \cellcolor{gray!15}.97/.07 & \cellcolor{gray!15}.44/.09 & \cellcolor{gray!15}.49/.14 & \cellcolor{gray!15}{\bf 1.00}/.06 & \cellcolor{gray!15}.47/.08 & \cellcolor{gray!15}.44/.10 & \cellcolor{gray!15}.95/.12 & \cellcolor{gray!15}.25/.04 & \cellcolor{gray!15}.42/.09 & \cellcolor{gray!15}.98/.08 & \cellcolor{gray!15}.20/.06 & \cellcolor{gray!15}.55/.13 & \cellcolor{gray!15}.90/{\bf .00} \\
 & \cellcolor{gray!15}B3 & \cellcolor{gray!15}.34/.05 & \cellcolor{gray!15}.32/.11 & \cellcolor{gray!15}.96/.09 & \cellcolor{gray!15}.43/.12 & \cellcolor{gray!15}.61/.15 & \cellcolor{gray!15}{\bf 1.00}/.07 & \cellcolor{gray!15}.47/.03 & \cellcolor{gray!15}.47/.15 & \cellcolor{gray!15}.95/.08 & \cellcolor{gray!15}.25/.08 & \cellcolor{gray!15}.45/.15 & \cellcolor{gray!15}.96/.08 & \cellcolor{gray!15}.25/.08 & \cellcolor{gray!15}.50/.09 & \cellcolor{gray!15}.90/{\bf .00} \\
\addlinespace[1pt]
\hline
\textit{S3 Path ID} & VERA & \multicolumn{3}{c|}{{\bf 1.00}/{\bf .00}} & \multicolumn{3}{c|}{{\bf 1.00}/{\bf .00}} & \multicolumn{3}{c|}{{\bf 1.00}/{\bf .00}} & \multicolumn{3}{c|}{{\bf 1.00}/{\bf .00}} & \multicolumn{3}{c}{{\bf 1.00}/{\bf .00}} \\
 & \cellcolor{gray!15}B3 & \cellcolor{gray!15}.68/{\bf .00} & \cellcolor{gray!15}.44/.14 & \cellcolor{gray!15}{\bf 1.00}/.10 & \cellcolor{gray!15}.75/.01 & \cellcolor{gray!15}.64/.12 & \cellcolor{gray!15}.99/.13 & \cellcolor{gray!15}.72/{\bf .00} & \cellcolor{gray!15}.52/.11 & \cellcolor{gray!15}{\bf 1.00}/.17 & \cellcolor{gray!15}.72/.03 & \cellcolor{gray!15}.52/.16 & \cellcolor{gray!15}{\bf 1.00}/.26 & \cellcolor{gray!15}.75/{\bf .00} & \cellcolor{gray!15}.42/.06 & \cellcolor{gray!15}{\bf 1.00}/.09 \\
\addlinespace[1pt]
\hline
\textit{S4 Reachability} & VERA & \multicolumn{3}{c|}{ .91/{\bf .00}} & \multicolumn{3}{c|}{.90/{\bf .00}} & \multicolumn{3}{c|}{.90/{\bf .00}} & \multicolumn{3}{c|}{.58/{\bf .00}} & \multicolumn{3}{c}{.44/{\bf .00}} \\
 &\cellcolor{gray!15} B3 & \cellcolor{gray!15}.35/{\bf .00} & \cellcolor{gray!15}.22/.01 & \cellcolor{gray!15}{\bf 1.00}/.22 & \cellcolor{gray!15}.20/{\bf .00} & \cellcolor{gray!15}.23/.05 & \cellcolor{gray!15}{\bf 1.00}/.15 & \cellcolor{gray!15}.33/{\bf .00} & \cellcolor{gray!15}.55/.02 & \cellcolor{gray!15}{\bf 1.00}/.22 & \cellcolor{gray!15}.26/{\bf .00} & \cellcolor{gray!15}.28/.01 & \cellcolor{gray!15}{\bf 1.00}/.15 & \cellcolor{gray!15}.28/{\bf .00} & \cellcolor{gray!15}.39/.01 & \cellcolor{gray!15}{\bf 1.00}/.16 \\
\addlinespace[1pt]
\hline
\textit{S5 Trace Cov.} & VERA & \multicolumn{3}{c|}{.59/{\bf .00}} & \multicolumn{3}{c|}{.64/{\bf .00}} & \multicolumn{3}{c|}{.71/{\bf .00}} & \multicolumn{3}{c|}{.62/{\bf .00}} & \multicolumn{3}{c}{.66/{\bf .00}} \\
 &\cellcolor{gray!15} B3 & \cellcolor{gray!15}.16/.02 & \cellcolor{gray!15}.73/.35 & \cellcolor{gray!15}{\bf 1.00}/{\bf .00} & \cellcolor{gray!15}.26/.07 & \cellcolor{gray!15}.79/.19 & \cellcolor{gray!15}{\bf 1.00}/{\bf .00} & \cellcolor{gray!15}.10/.03 & \cellcolor{gray!15}.69/.29 & \cellcolor{gray!15}{\bf 1.00}/{\bf .00} & \cellcolor{gray!15}.29/.11 & \cellcolor{gray!15}.63/.26 & \cellcolor{gray!15}{\bf 1.00}/{\bf .00} & \cellcolor{gray!15}.44/.08 & \cellcolor{gray!15}.69/.30 & \cellcolor{gray!15}.94/.01 \\
\addlinespace[1pt]
\hline
\textit{S6 State Trans.} & VERA & \multicolumn{3}{c|}{{\bf 1.00}/.01} & \multicolumn{3}{c|}{{\bf 1.00}/.02} & \multicolumn{3}{c|}{{\bf 1.00}/.02} & \multicolumn{3}{c|}{{\bf 1.00}/.02} & \multicolumn{3}{c}{{\bf 1.00}/.05} \\
 & \cellcolor{gray!15}B3 & \cellcolor{gray!15}.00/.09 & \cellcolor{gray!15}.62/.26 & \cellcolor{gray!15}{\bf 1.00}/{\bf .00} & \cellcolor{gray!15}.20/.08 & \cellcolor{gray!15}.90/.21 & \cellcolor{gray!15}{\bf 1.00}/{\bf .00} & \cellcolor{gray!15}.42/.16 & \cellcolor{gray!15}.75/.32 & \cellcolor{gray!15}{\bf 1.00}/{\bf .00} & \cellcolor{gray!15}.45/.05 & \cellcolor{gray!15}.91/.12 & \cellcolor{gray!15}{\bf 1.00}/{\bf .00} & \cellcolor{gray!15}.50/.07 & \cellcolor{gray!15}.75/.38 & \cellcolor{gray!15}{\bf 1.00}/.01 \\
\addlinespace[1pt]
\hline
\textit{S7 Alias} & VERA & \cellcolor{gray!15}.17/.03 & \cellcolor{gray!15}.17/.20 & \cellcolor{gray!15}.50/.03 & \cellcolor{gray!15}.33/.06 & \cellcolor{gray!15}.75/.16 & \cellcolor{gray!15}{\bf 1.00}/.04 & \cellcolor{gray!15}.25/.08 & \cellcolor{gray!15}.50/.13 & \cellcolor{gray!15}{\bf 1.00}/.06 & \cellcolor{gray!15}.33/.06 & \cellcolor{gray!15}.83/.25 & \cellcolor{gray!15}{\bf 1.00}/.05 & \cellcolor{gray!15}{\color{gray}—}/.06 & \cellcolor{gray!15}{\color{gray}—}/.15 & \cellcolor{gray!15}{\color{gray}—}/.01 \\
 & \cellcolor{gray!15}B3 & \cellcolor{gray!15}.17/.06 & \cellcolor{gray!15}.17/.27 & \cellcolor{gray!15}.50/.03 & \cellcolor{gray!15}.25/.03 & \cellcolor{gray!15}.50/.22 & \cellcolor{gray!15}{\bf 1.00}/.02 & \cellcolor{gray!15}.00/.09 & \cellcolor{gray!15}.75/.25 & \cellcolor{gray!15}{\bf 1.00}/.06 & \cellcolor{gray!15}.33/.07 & \cellcolor{gray!15}.67/.10 & \cellcolor{gray!15}{\bf 1.00}/.05 & \cellcolor{gray!15}{\color{gray}—}/.02 & \cellcolor{gray!15}{\color{gray}—}/.26 & \cellcolor{gray!15}{\color{gray}—}/.01 \\
\addlinespace[1pt]
\hline
\textit{S8 Verdict} & VERA & \multicolumn{3}{c|}{{\bf 1.00}/{\bf .00}} & \multicolumn{3}{c|}{{\bf 1.00}/{\bf .00}} & \multicolumn{3}{c|}{{\bf 1.00}/{\bf .00}} &\multicolumn{3}{c|}{{\bf 1.00}/.02} & \multicolumn{3}{c}{{\bf 1.00}/{\bf .00}} \\
 &\cellcolor{gray!15} B3 & \cellcolor{gray!15}{\bf 1.00}/{\bf .00} & \cellcolor{gray!15}.75/.09 & \cellcolor{gray!15}{\bf 1.00}/.19 & \cellcolor{gray!15}.98/{\bf .00} & \cellcolor{gray!15}.85/.10 & \cellcolor{gray!15}.98/.19 & \cellcolor{gray!15}{\bf 1.00}/{\bf .00} & \cellcolor{gray!15}.94/.09 & \cellcolor{gray!15}{\bf 1.00}/.19 & \cellcolor{gray!15}.86/{\bf .00} & \cellcolor{gray!15}.80/.09 & \cellcolor{gray!15}.89/.18 & \cellcolor{gray!15}.89/{\bf .00} & \cellcolor{gray!15}.78/.04 & \cellcolor{gray!15}{\bf 1.00}/.14 \\
\bottomrule
\end{tabular}}

\end{table*}

\subsection{RQ2: Judge Effectiveness} 
To answer RQ2, we randomly sampled 20 SRRs per dataset per CWE, applied each mutation operator (prioritizing \texttt{No\_ERROR}, falling back to \texttt{MINOR\_ERROR}), and measured per-stage recall and false-positive (FP) rates across the resulting mutated records. When scoring, pre-existing errors in the unmutated base records were excluded, ensuring measured recall reflects detection of the injected fault.
For each mutant, we checked whether the judge flagged the specific stage targeted by the operator. We compared VERA against B3 (SRR + structured LLM judge), its LLM-only counterpart, which audits SRRs across the same eight stages but uses an LLM for every stage. Tbl.~\ref{tab:rq2} reports per-stage recall and off-target FP for both judges across three judge models, aggregating records of the same CWE across datasets; all SRRs are generated by Sonnet-5, which produces the most structurally complete records and all judges are evaluated on the identical set of mutants.

\boldparagraph{Primarily deterministic stages (S1, S3--S6, S8)} VERA matches or outperforms B3 (SRR + structured LLM judge) across all these stages while maintaining near-zero off-target FP. S3, S6, and S8 achieve perfect recall ($1.00$) with $0.00$ FP across all CWEs. Across these stages, B3 with GPT-4.1 as judge achieves recall 0.10--0.75 with FP up to 0.16; B3 with DeepSeek achieves 0.22--0.91 in recall but at FP up to 0.38 on clean records. B3 with Mistral reaches near-perfect recall on most of these stages, but only because of overflagging incurring FP up to 0.26. 
The key pattern underlying VERA's advantage is \emph{absence detection}: the errors targeted by S5 (missing intermediate state changes) and S6 (omitted or misordered transitions) require noticing what is not in the trace rather than what is. Fluent surrounding prose masks these gaps for an LLM judge — VERA's deterministic trace checks find them mechanically. S5 shows the largest gap overall: VERA 91.5\% vs.\ B3 72.8\%, with 22.8\% of injected errors caught by VERA alone. S6 shows the same pattern (+16.7\%). The one exception to VERA's structural advantage is S1, where recall falls slightly short due to a deliberate precision-first trade-off (Sec.~\ref{sec:judge}); even so, S1 matches B3 recall with lower FP on two of three judge models. Notably, S8 (verdict consistency) is the one stage where the gap nearly closes regardless of judge type — VERA 100\%, B3 98.8\% — because an explicit contradiction between the property block and the final verdict is detectable in any structured representation without absence reasoning.
Beyond quantitative metrics, B3 (SRR + structured LLM) exhibits severe hallucination on clean records. With GPT-4.1, on the \texttt{SMB2\_read} control record (Linux kernel, CWE-416 in SVEN), B3 issued a \textsc{Fail} on S3 with the finding: \emph{``Hallucinated State Transition: rsp and *buf step 12, 15.''} No such transition exists in either the record or the source — B3 invented a violation on a sound reasoning chain. VERA correctly returned \textsc{Pass}.

\boldparagraph{Primarily LLM-dependent stages (S2, S7)} Neither VERA nor B3 (SRR + structured LLM) achieves reliable detection on S2 and S7, and performance differences between them are attributable to LLM stochasticity rather than systematic design advantages. S2 (assumptions) recall reaches 0.20--0.47: drawing a strict line between grounded and ungrounded claims is inherently ambiguous, and the gap between VERA and B3 here falls within stochastic variance. S7 (alias) exhibits the lowest recall overall, as alias detection requires comprehensive points-to analysis over the global pointer graph, which static analysis alone cannot perform on localized code slices. These results reflect the current limitations of LLM semantic reasoning on these two error classes. We retain S2 and S7 because the failure modes they target are prevalent and consequential; closing the recall gap on both remains an active direction for future work.

\boldparagraph{Model variance and the precision--recall trade-off} VERA's programmatic stages (S1, S3--S6, S8) are fully deterministic and identical across all three judge models. Variance appears only on S2 and S7. Mistral reaches near-perfect recall on several stages but with FP up to 0.26, indicating instability; DeepSeek is more conservative but has higher FP than GPT-4.1. This exposes another fundamental limitation of a purely LLM-based judge: flagging more aggressively catches more injected errors but also raises more false positives, while being conservative reduces false positives at the cost of missing real errors. VERA's programmatic stages achieve higher recall than the conservative GPT judge, and lower FP than the aggressive Mistral judge, avoiding both extremes.


In summary, across five CWE classes and three state-of-the-art judge models, VERA matches or outperforms the structured LLM judge (B3) on every programmatic stage while keeping off-target FP at near zero on three of these stages. Thus, we answer RQ2.

\begin{table*}[t]
\centering
\caption{RQ3 comparing VERA structured framework with free-form baselines. }
\label{tab:rq3_combined}
\footnotesize\setlength{\tabcolsep}{3pt}
\subcaptionbox{SRR construction success rate per detection model. \label{tab:rq3_a}}[0.27\linewidth]{%
\centering
\scalebox{0.8}{
\begin{tabular}{l rrrr}
\toprule
\textbf{Model} & \textbf{Total}  & \textbf{Incomp.} & \textbf{success\%} \\
\midrule
\textit{Sonnet-5} & 1649& 46 & 97.2\% \\
\textit{Gemma} & 1654  & 6 & 99.6\% \\
\textit{GLM} & 1632  & 6 & 99.6\% \\
\textit{Llama} & 1652 & 37 & 97.8\% \\
\textit{Nemotron} & 1564 & 398 & {\bf 74.6\%} \\
\textit{Qwen} & 1646 & 25 & 98.5\% \\
\bottomrule
\end{tabular}}
}%
\hfill
\subcaptionbox{Failure rate flagged by VERA, B1 and B2 for the determinisitic stages. \label{tab:rq3_b}}[0.4\linewidth]{%
\centering
\scalebox{0.8}{
\begin{tabular}{l | c | c c c | c c c}
\toprule
\textbf{Stage} & \textsc{vera} & \multicolumn{3}{c|}{\textit{B1 }} & \multicolumn{3}{c}{\textit{B2 }} \\
 & & \textit{GPT} & \textit{DS} & \textit{Mis} & \textit{GPT} & \textit{DS} & \textit{Mis} \\
\midrule
S1 Scope & 59\% & 4\% & 4\% & 4\% & 29\% & 64\% & 51\% \\
\addlinespace[2pt]
S3 Path ID & 17\% & 3\% & 3\% & 3\% & 11\% & 70\% & 66\% \\
\addlinespace[2pt]
S4 Reach. & 0\% & 0\% & 0\% & 0\% & 0\% & 1\% & 0\% \\
\addlinespace[2pt]
S5 Trace Cov. & 26\% & 10\% & 10\% & 11\% & 76\% & 76\% & 97\% \\
\addlinespace[2pt]
S6 State Trans. & 42\% & 5\% & 5\% & 6\% & 13\% & 49\% & 66\% \\
\addlinespace[2pt]
S8 Verdict & 65\% & 0\% & 0\% & 0\% & 0\% & 5\% & 2\% \\
\bottomrule
\end{tabular}}
}%
\hfill
\subcaptionbox{Pairwise agreement between VERA, B1, and B2.\label{tab:rq3_c}}[0.28\linewidth]{%
\centering
\scalebox{0.65}{
\begin{tabular}{l l | r r r}
\toprule
\multicolumn{2}{l}{\textbf{Pair}} & \textbf{Both N.E} & \textbf{Both E.} & \textbf{Disagree} \\
\midrule
\multirow{3}{*}{\shortstack{VERA \\vs. B1}} & \textit{GPT} & 6\% & 32\% & 61\% \\
 & \textit{DS} & 7\% & 20\% & 72\% \\
 & \textit{Mis} & 5\% & 38\% & 55\% \\
\hline
\multirow{3}{*}{\shortstack{VERA \\vs. B2}} & \textit{GPT} & 0\% & 78\% & 20\% \\
 & \textit{DS} & 0\% & 89\% & 9\% \\
 & \textit{Mis} & 0\% & 91\% & 8\% \\
\hline
\multirow{3}{*}{B1 vs.\ B2} & \textit{GPT} & 10\% & 31\% & 58\% \\
 & \textit{DS} & 1\% & 21\% & 76\% \\
 & \textit{Mis} & 0\% & 41\% & 57\% \\
\midrule
\multicolumn{5}{l}{\textit{Deterministic stages only. N.E.: no error; E.: error}}\\
\bottomrule
\end{tabular}}
}
\\[5pt]
\subcaptionbox{Stage-level error rates split by final verdict correctness and ground-truth label.\label{tab:rq3_d}}[0.46\linewidth]{%
\centering
\scalebox{0.72}{
\begin{tabular}{l | r r |r  r r}
\toprule
\textbf{Stage} & \multicolumn{2}{c|}{\textit{(i) Verdict Correctness}}  & \multicolumn{3}{c}{\textit{(ii) Ground-truth }} \\
 & \textbf{Corr.} & \textbf{Incorr.} & \textbf{Vuln} & \textbf{Safe}  & \textbf{Both Same Err}\\
\midrule
S1 Scope & .51 & .48  & .49 & .51 & .44  \\
S3 Path ID & .23 & .22  & .22 & .23 & .13  \\
S4 Reach. & .00 & .00  & .00 & .00  & .00  \\
S5 Trace Cov. & .31 & .31  &  .31 & .31  & .15 \\
S6 State Trans. & .39 & .39  &  .38 & .39  & .23 \\
S8 Verdict & .64 & .59  &  .61 & .62 & .50  \\

\bottomrule
\end{tabular}
}}
\hfill
\subcaptionbox{Stage-level failure rate per detection model. \label{tab:rq3_e}}[0.52\linewidth]{%
\centering
\scalebox{0.8}{
\begin{tabular}{l | cccccc}
\toprule
\textbf{Stage} & \textit{Sonnet-5} & \textit{Gemma} & \textit{GLM} & \textit{Llama} & \textit{Nemotron} & \textit{Qwen} \\
\midrule
S1 Scope & \cellcolor{red!35}.59 & \cellcolor{orange!30}.48 & \cellcolor{red!35}.51 & \cellcolor{orange!30}.48 & \cellcolor{orange!30}.47 & \cellcolor{orange!30}.48 \\
S3 Path ID & \cellcolor{yellow!20}.18 & \cellcolor{orange!30}.34 & \cellcolor{yellow!20}.19 & \cellcolor{yellow!20}.26 & \cellcolor{yellow!20}.21 & \cellcolor{yellow!20}.13 \\
S4 Reach. & .00 & .00 & .00 & .00 & .00 & .00 \\
S5 Trace Cov. & \cellcolor{yellow!20}.22 & \cellcolor{orange!30}.33 & \cellcolor{yellow!20}.27 & \cellcolor{yellow!20}.26 & \cellcolor{yellow!20}.11 & \cellcolor{yellow!20}.16 \\
S6 State Trans. & \cellcolor{orange!30}.43 & \cellcolor{yellow!20}.23 & \cellcolor{orange!30}.42 & \cellcolor{orange!30}.44 & \cellcolor{orange!30}.32 & \cellcolor{orange!30}.47 \\
S8 Verdict & \cellcolor{red!35}.66 & \cellcolor{red!35}.50 & \cellcolor{red!35}.69 & \cellcolor{red!35}.54 & \cellcolor{red!35}.51 & \cellcolor{red!35}.67 \\
\bottomrule
\end{tabular}}
}%

\end{table*}

\subsection{RQ3: LLM Reasoning Quality Information Revealed} 
To answer \textbf{RQ3}, we compare VERA's performance on its deterministic stages against free-form baselines B1 (CoT + Holistic LLM) and B2 (CoT + Structured LLM). We focus on stages that reached decent performance in RQ2, excluding S2 and S7. 
First, we evaluate whether detection models can reliably output SRRs. Tbl.~\ref{tab:rq3_a} shows that five of the six detection models generate well-formed SRRs in $\ge 97.2\%$ of cases; only Nemotron fails to produce a parseable record on $25.4\%$ of attempts. 
For Nemotron, the incomplete records are excluded from all downstream stage-level analysis.

\boldparagraph{Free-form baselines vs.\ VERA} Overall, B1 (CoT + Holistic LLM) yields an $86.8\%$ false-negative rate on records where VERA identifies deterministic-stage errors, underscoring the inadequacy of free-form LLM evaluation. As shown in Tbl.~\ref{tab:rq3_b}, B1 (CoT + Holistic LLM) flags substantially fewer errors than VERA across every stage presented, whereas B2 (CoT + Structured LLM) over-flags in almost all stages while missing errors in S8. Discrepancies are largest on stages requiring precise cross-section consistency (S3, S5, S6, S8). A representative example is \texttt{disk\_seqf\_stop} (SVEN CWE-416, Sonnet-5): the SRR explicitly assigns \texttt{iter} as \texttt{NULL} then dereferences it in the next step, yet marks \texttt{CWE-476} as \textsc{Satisfied}. VERA's S8 detects this contradiction mechanically; B1 (CoT + Holistic LLM) rates the same SRR error-free, stating \emph{``The analysis correctly describes the bug, tracing all relevant pointer/state paths.''} A holistic LLM judge infers coherence from parts that sound correct rather than verifying mutual consistency — which is precisely what structured stage checks provide. 
%
LLM-as-a-judge is inherently prompt-sensitive and therefore unreliable~\cite{ZhuWW25}. As shown in Tbl.~\ref{tab:rq3_c}, B1 (CoT + Holistic LLM) and B2 (CoT + Structured LLM) disagree significantly on the same CoT inputs, confirming that free-form judging lacks the consistency needed for reliable reasoning auditing.


\boldparagraph{Finding 1: Correct verdicts are no guarantee of correct reasoning} Tbl.~\ref{tab:rq3_d} partitions audited SRRs by verdict correctness. Intuitively, correct verdicts should show fewer reasoning flaws — but error rates across all six stages remain virtually identical between correct and incorrect verdicts ($|\delta| \leq 0.05$ at every stage). Models are frequently \emph{right for the wrong reasons}: even when they output correct binary verdicts, their reasoning contains explicit internal contradictions invisible to surface-level accuracy metrics. A similar decoupling holds across the ground-truth split, with error rates uniform across safe and vulnerable code.

\boldparagraph{Finding 2: LLM over-flag by pattern-matching, not by reasoning} Among 1,468 functions evaluated by all six models, 103 safe functions were wrongly labelled as vulnerable by every model, compared to only 37 real vulnerabilities missed by all models. VERA's structured audit flagged a reasoning error in $98\%$ of these over-flagged safe functions, with S8 as the dominant failure ($82\%$). This reflects a reliance on surface pattern matching observed in Sec.~\ref{sec:motivation}, models flag safe functions for containing vulnerable-like structural cues, then hallucinate ungrounded traces to justify their premature \textsc{Vulnerable} verdict.
Examining paired data confirms this. Across 777 paired (vulnerable, patched) functions, both versions trigger the same stage failure in $51.1\%$ of pairs at S1, $27.7\%$ at S6, and $56.9\%$ at S8 — and $32.3\%$ of pairs share the exact errors across all stages. The model makes the same structural reasoning mistakes regardless of whether the vulnerability is present, confirming that over-prediction is driven by code pattern 
rather than vulnerability-specific reasoning.

\boldparagraph{Finding 3: Incomplete slices cause models to hallucinate lifecycle evidence} 
While over-flagging and miss rates are comparable for most CWEs, CWE-772 (Missing Release) is a major outlier: models correctly flag $75\%$ of actual leaks (highest recall across CWEs), yet falsely flag $73\%$ of safe functions. This stems from a mismatch between CWE-772 violation and slice visibility. Unlike local properties like CWE-476, confirming a CWE-772 leak requires verifying that the resource reaches a release state on \emph{every} exit path — including paths that continue beyond the slice boundary. When an explicit release is absent from the local slice, models tend to default to predicting a leak. VERA's stage results makes this concrete: S5 (Trace Coverage) fails at $30.8\%$ for CWE-772 (vs.\ $17.6\%$ for CWE-476). Furthermore, S6 (State Transitions) fails at $58.2\%$ (vs.\ $38.6\%$ for CWE-476), due to hallucinated release operations (\texttt{nullify(image)} claimed in the trace but absent from source) and misordered transitions. 
Across both stages, models exhibit the same failure mode: when crucial lifecycle context lies beyond the visible slice, models commit prematurely to a verdict and insert unsupported trace artifacts to bridge the gap. VERA detects these as code adherence failures and flags them for human follow-up. An analyst can then inspect the full codebase to determine whether the claimed operations exist outside the slice boundary.


\boldparagraph{Finding 4: Code patterns per stage} 
By decomposing reasoning chains into distinct stages, VERA reveals how specific code constructs systematically induce distinct failure modes. 
Stage $S_5$ (Trace Coverage) failures concentrate on structurally complex functions: failed cases exhibit $1.66\times$ more loops, $1.52\times$ more null checks, and $0.49\times$ as many allocation calls compared to functions without reasoning errors (all $p < 0.01$, Mann-Whitney $U$~\cite{mann1947test}). 
In contrast, $S_6$ (State Transitions) failures show no significant correlation with structural complexity, including loops, branches, and free calls ($p > 0.05$ for all). Finally, $S_8$ (Verdict Consistency) failures occur significantly more often in functions with frequent null checks ($1.61\times$, $p < 0.001$) and fewer allocation calls ($0.63\times$, $p < 0.001$), whereas explicit null-assignment patterns (\texttt{= NULL}) mitigate verdict inconsistency ($0.54\times$, $p < 0.001$).

\boldparagraph{Finding 5: Different LLMs make different reasoning mistakes} VERA surfaces distinct error profiles across detection models, offering actionable results for model selection. The most universal finding is S8: every model fails this stage at $50-69\%$, revealing that reconciling reasoning with the final verdict is a systematic weakness across all current architectures — and one that B1 misses almost entirely (Tbl.~\ref{tab:rq3_b}). Beyond S8, models diverge: Sonnet-5 and Qwen achieve the lowest S3 and S5 error rates ($13-22\%$) while Gemma struggles ($33-34\%$); yet Gemma leads on S6 ($23\%$) while Qwen and Sonnet-5 incur $43-47\%$. This heterogeneity enables targeted refinement — identifying a model's exact failure stage rather than optimizing blindly for binary accuracy — and principled selection: models stronger on path traversal (Sonnet-5, Qwen) for deep control-flow vulnerabilities; models stronger on local state transitions (Gemma) for use-after-free lifecycles.

In summary, VERA reveals that reasoning quality is not related to binary accuracy, uncovers cross-section inconsistencies and systematic over-prediction that free-form judges miss, and provides fine-grained model diagnostic profiles invisible to standard benchmarks. Thus, we answer RQ3.

\subsection{Threats to Validity}
1) \textbf{Construct — training data contamination.} Public benchmarks like SVEN and PrimeVul may have been exposed to LLMs during pre-training.
Contamination could bias models toward memorized binary verdicts rather than genuine code analysis. We mitigated this with extra CVE data collected after the model training date, and observed no significant difference in reasoning error rates, suggesting contamination alone does not explain the failures observed.
2) \textbf{External — CWE scope.} Our evaluation covers five C/C++ memory and resource lifecycle CWEs, which are among the most exploited and hardest to detect vulnerability classes in practice. Extending VERA to other CWEs is left as future work.
3) \textbf{Internal — structured format bias.} VERA can only check what is explicitly recorded in the SRR. A model that produces sparse or underspecified output gives VERA fewer opportunities to detect errors, which may inflate its sound rate independent of actual reasoning quality. We observed this with smaller open-source models; in this study, the five frontier models we evaluate reliably adhere to the schema. 
4) \textbf{Internal- Mutation-judge Alignment.} A potential concern is that our mutation operators ($M_1\text{--}M_{8}$) target structural properties inherently favored by VERA's deterministic checks over LLM judges. We mitigate this by grounding the operators in naturally occurring errors. RQ1's open-ended annotation confirms that $M_1$–$M_8$ directly instantiate the error categories ($F_1$–$F_8$) that arise in real, un-mutated LLM reasoning. The mutations are therefore representative of real failures, not constructions designed to favor VERA.

\section{Related Work}
\label{sec:related_work}

\boldparagraph{LLM-based vulnerability detection} Learning-based vulnerability detection has moved from neural models over program representations~\cite{Zhou2019Devign,chakraborty2022reveal,DARPA2026,du2026vul,su2024codeart,du2024generalization} to large language models fine-tuned on code~\cite{zhou2025llmvulsurvey}, with recent systems adding knowledge retrieval~\cite{Du2024VulRAG}, code-graph guidance~\cite{lekssays2025llmxcpg}, or taint inference~\cite{li2025iris}. These systems optimize for detection accuracy; VERA is not a detector, it audits the reasoning that a detection model records in an SRR.

\boldparagraph{Validating LLM-found vulnerabilities} LLM-based vulnerability detection increasingly include validity check of their vulnerability reports. LLift~\cite{li2024llift} prompts an LLM to review its responses against a set of rules, LLMSAN~\cite{wang2024llmsan} discards reports whose recorded data-flow paths fail its checks; LLMDFA~\cite{wang2024llmdfa} checks path feasibility with an SMT solver and RepoAudit~\cite{guo2025repoaudit} asks the LLM to find contradictions among path conditions; BugScope~\cite{guo2026bugscope} checks feasibility and whether its assumptions are justified by the code solely through an LLM-as-judge; VulInstruct~\cite{zhu2026vulinstruct} and DAGVul~\cite{lu2026dagvul} check for correctness by comparing with ground truth information also with LLM-as-judge. These approaches ask whether a reported vulnerability is real; VERA checks the reasoning behind the verdict. These two goals are complementary to each other, a validator can confirm a bug exists; VERA can reveal whether the LLM found it for the right reasons.  

\boldparagraph{Checking LLM reasoning} Prior work has examined the reasoning behind vulnerability verdicts, but all rely on free-form CoT and LLM-as-judge. 
Free-form CoT has been shown to be unreliable, including in general natural language tasks~\cite{turpin2023unfaithful,arcuschin2026wild} and specifically for code reasoning~\cite{ullah2024secllmholmes,DARPA2026,li2025svtrusteval}. LLM judges are widely used to evaluate model outputs~\cite{Zheng2023JudgingLLM,crupi2025judge,wang2025replacehuman}, including for code generation and SE tasks~\cite{chen2025reval,liu2026codemind,abdollahi2026demystifying,crupi2025judge,wang2025replacehuman}, despite being susceptible to superficial cues~\cite{moon2026cover}.
Some vulnerability detection approaches attempt to check reasoning correctness, as mentioned before, VulInstruct~\cite{zhu2026vulinstruct} and DAGVul~\cite{lu2026dagvul}, but do so only by comparing generated reasoning against ground truth obtained through patches or human annotation using a holistic LLM-as-judge. VERA instead checks claimed scope, state transitions, and assumptions directly against the code slice, requiring no ground-truth annotations and enabling deterministic checks of a few reasoning failures.

\boldparagraph{Evaluating reasoning judges}
Most prior work evaluates vulnerability detectors against labeled ground-truth datasets. However, no such labels exist for reasoning quality, making judge evaluation significantly harder.
Prior work evaluates judges on non-coding tasks by introducing predefined quality mutations into model outputs: FBI~\cite{doddapaneni2024fbi} mutates natural language output quality (e.g., coherence and fluency) against checklist criteria, and Delcourt et al.~\cite{delcourt2026breaking} mutate UML diagrams by injecting structural defects. This does not directly transfer to code reasoning: a meaningful mutation must identify a specific claim to modify, ensure it is wrong with respect to the source code, and attach a known failure category, but free-form CoT provides no clear anchors for any of these. VERA enables this by combining the SRR, which provides explicit claim boundaries and source grounding, with the failure taxonomy that defines typed failure categories to mutate.

\section{Conclusion and Future Work}
\label{sec:conclusion}

This paper presented VERA, a framework for automated auditing of LLM vulnerability reasoning that replaces free-form LLM-as-judges with a deterministic/LLM pipeline over machine-readable Structured Reasoning Records. Our evaluation shows that reasoning quality is 
decoupled from verdict correctness: LLMs are frequently right for the wrong reasons, and this goes undetected by standard accuracy metrics and pure LLM-as-judges alike — LLM-as-judges miss $87\%$ of the reasoning flaws VERA detects mechanically. VERA's per-stage decomposition further reveals structural correlates of reasoning failures invisible to binary accuracy: loop-heavy functions concentrate trace coverage errors, while null-heavy pointer patterns drive verdict inconsistencies. 
In practice, a security analyst using an LLM-based vulnerability scanner can run VERA to filter hallucinated reports before acting on them, while LLM developers can use its stage-wise failure profiles as targeted fine-tuning signals.
Future work will extend VERA beyond C/C++ memory vulnerabilities to semantic flaw classes, integrate interprocedural static analysis to improve pointer alias recall, and leverage VERA's stage-wise diagnostics as fine-grained training signals for LLM alignment.

\section{Data Availability}
Our implementation of VERA, additional results and supplementary material can be found at \url{https://anonymous.4open.science/r/VERA-D35F}.

\bibliographystyle{ACM-Reference-Format}
\bibliography{ref}




\end{document}